%% file: SUS-12-002_temp.tex
\pdfoutput=1

\documentclass[11pt,twoside,a4paper,cmspaper,final,collab]{cms-tdr}
\def\svnVersion{135292:135650}\def\cmsCernNo{2012-186}\def\cmsCernDate{\today}\def\cmsMessage{Submitted to the Journal of High Energy Physics}

\begin{document}\cmsNoteHeader{SUS-12-002}

\hyphenation{had-ron-i-za-tion}
\hyphenation{cal-or-i-me-ter}
\hyphenation{de-vices}

\RCS$Revision: 135650 $
\RCS$HeadURL: svn+ssh://alverson@svn.cern.ch/reps/tdr2/papers/SUS-12-002/trunk/SUS-12-002.tex $
\RCS$Id: SUS-12-002.tex 135650 2012-07-07 14:19:38Z alverson $
\newcommand{\Lumino}{4.73\fbinv}
\newcommand{\MT}{\ensuremath{M_{\mathrm{T}}}\xspace}
\newcommand{\MTtwo}{\ensuremath{M_{\mathrm{T2}}}\xspace}
\newcommand{\MTtwob}{\ensuremath{M_{\mathrm{T2}}b}\xspace}
\newcommand{\METvec}{\ensuremath{\vec{E}_{\mathrm{T}}^{\text{miss}}}\xspace}
\newcommand{\MHT}{\ensuremath{H_{\mathrm{T}}^{\text{miss}}}\xspace}
\newcommand{\MHTvec}{\ensuremath{\vec{H}_{\mathrm{T}}^{\text{miss}}}\xspace}
\newcommand{\ptmissvec}{\ensuremath{\vec{p}_{\mathrm{T}}^{\; \text{miss}}}\xspace}
\newcommand{\ptvec}{\ensuremath{\vec{p}_{\mathrm{T}}}\xspace}
\providecommand{\PSGc}{\ensuremath{\widetilde{\chi}}\xspace} % neutralino
\cmsNoteHeader{SUS-12-002} % This is over-written in the CMS environment: useful as preprint no. for export versions
\title{Search for supersymmetry in hadronic final states using \texorpdfstring{\ensuremath{M_{\mathrm{T2}}} in pp collisions at $\sqrt{s}$ = 7\TeV}{MT2 in pp collisions at sqrt(s)=7 TeV}}

\date{\today}

\abstract{
A search for supersymmetry or other new physics resulting in similar final
states is presented using
a data sample of 4.73\fbinv of pp
collisions collected at
$\sqrt{s}=7\TeV$
with the CMS detector at the LHC.
Fully hadronic final states are selected based on the variable
\ensuremath{M_{\mathrm{T2}}}, an
extension of the transverse mass in events with two invisible particles.
Two complementary studies are performed.
The first targets the region of parameter space with medium to high squark and
gluino masses,
in which the signal can be separated from the standard model backgrounds by a
tight requirement on \ensuremath{M_{\mathrm{T2}}}.
The second is optimized to be sensitive to events with a light gluino and
heavy squarks.
In this case, the \ensuremath{M_{\mathrm{T2}}} requirement is relaxed,
but a higher jet multiplicity
and at least one b-tagged jet are required.
No significant excess of events over the standard model expectations is
observed. Exclusion limits are derived for the parameter space of the
constrained minimal supersymmetric extension of the standard model, as well as
on a variety of simplified model spectra.
}

\hypersetup{%
pdfauthor={CMS Collaboration},%
pdftitle={Search for supersymmetry in hadronic final states using MT2 in pp collisions at sqrt(s) = 7 TeV},%
pdfsubject={CMS},%
pdfkeywords={CMS, physics, supersymmetry}}

\maketitle %maketitle comes after all the front information has been supplied

\input{PAPER_MT2_core}

\bibliography{auto_generated}   % will be created by the tdr script.
\cleardoublepage \appendix\section{The CMS Collaboration \label{app:collab}}\begin{sloppypar}\hyphenpenalty=5000\widowpenalty=500\clubpenalty=5000\input{SUS-12-002-authorlist.tex}\end{sloppypar}
\end{document}

%% file: PAPER_MT2_core.tex
\section{Introduction}

A broad class of extensions of the standard model (SM) predict the existence of heavy colored particles
that decay to hadronic final states accompanied by large missing transverse energy (\MET).
The best known of these scenarios is supersymmetry~\cite{Martin:1997ns} (SUSY) with R-parity conservation.
In this paper we present a search for such new physics in pp collisions collected with the Compact Muon Solenoid (CMS) detector at the Large Hadron Collider (LHC)
at a center-of-mass energy of 7\TeV. The results are based on the data sample collected in 2011,
corresponding to about \Lumino of integrated luminosity.

The search makes use of the ``stransverse mass'' variable \MTtwo \cite{Lester:1999tx, Barr:2003rg}
to select new physics candidate events.
\MTtwo is the natural extension of the transverse mass \MT to the case where two colored supersymmetric particles (``sparticles'') are pair-produced and both decay through a cascade
of jets and possibly leptons to the lightest supersymmetric particle (LSP).
The LSP is not visible in the detector and leads to a missing transverse momentum signature.
Although \MTtwo was originally introduced to derive the masses of sparticles involved in the cascade decay, we use it here
as a discovery variable since it is sensitive to the presence of SUSY-like new physics.
The distribution of \MTtwo reflects the produced particle masses,
which are much lighter for the SM background processes
than for the SUSY processes.
Hence, new physics is expected to appear as an excess in the tail of \MTtwo.

The analysis is based on two complementary approaches.
A first approach, the ``\MTtwo analysis'', targets events resulting from heavy sparticle production,
characterized by large \MET, at least three jets, and large \MTtwo.
The SM backgrounds in the signal region
consist of W$(\ell\nu)$+jets, Z$(\cPgn\cPagn)$+jets, \ttbar,
and single-top events
(the last two will be referred to collectively as top-quark background),
which are estimated from data-control regions and simulation.
This analysis loses sensitivity if the squarks are heavy and the gluinos light,
in which case the production is dominated by gluino-gluino processes.
The gluinos give rise to three-body decays with relatively small \MET.
Since the gluino decay is mediated by virtual squark exchange
and the stop and sbottom are expected to be lighter than the first- and second-generation squarks,
these events can be rich in b quarks.
To increase the sensitivity to such processes,
a second approach, the ``\MTtwob analysis'', is developed,
in which the threshold on \MTtwo defining the signal region is lowered.
To suppress the QCD multijet background, we demand at least
one b-tagged jet and place a stricter requirement on the jet multiplicity.
The \MTtwob analysis provides a larger signal-to-background ratio
in the region of heavy squarks and light gluinos
and hence improves our sensitivity to this scenario.

This paper extends previous results of searches in fully hadronic final states
from the CMS~\cite{RA2paper35pb-1,RAZORpaper35pb-1,AlphaTPaper35pb-1,AlphaTPaper1fb-1}
and ATLAS~\cite{Aad201267,ATLASJetMET,ATLAS2011,Aad:2012hm} Collaborations.
It is organized as follows:
after a brief introduction to \MTtwo and its salient properties in Section~\ref{sec:MT2},
and a description of the CMS detector in Section~\ref{sec:CMS},
we present in Section~\ref{sec:cut-flow} the data samples used and the event selection.
In Section~\ref{sec:strategy}, the search strategy is presented.
This strategy is applied to the \MTtwo analysis in Section~\ref{sec:data:geq3jets}
and to the \MTtwob analysis in Section~\ref{sec:data:relaxed}.
In these sections the background estimation methods are also discussed.
We interpret the results in Section~\ref{sec:exclusion} in the context of the
constrained minimal supersymmetric standard model (CMSSM) as well as for a
variety of simplified models.
Finally, Section~\ref{sec:conclusion} contains a summary.

\section{Definition of \texorpdfstring{\MTtwo}{MT2}}
\label{sec:MT2}

The variable \MTtwo was introduced \cite{Lester:1999tx}
to measure the mass of primary pair-produced particles
in a situation where both ultimately decay into undetected particles
(e.g., LSPs) leaving the event kinematics underconstrained.
It assumes that the two produced sparticles give rise to identical types of decay chains
with two visible systems defined by their transverse momenta $\vec{p}_{\mathrm{T}}^{\; \mathrm{vis}(i)}$,
transverse energies $E_{\mathrm{T}}^{\mathrm{vis}(i)}$, and masses $m^{\mathrm{vis}(i)}$.
They are accompanied by the unknown LSP transverse momenta $\vec{p}_{\mathrm{T}}^{\; \PSGc{(i)}}$.
In analogy with the transverse mass used for the W boson mass determination~\cite{Arnison1983103},
we can define two transverse masses ($i = 1, 2$):
\begin{eqnarray}
	(M_{\mathrm{T}}^{{(i)}})^2 = (m^{\text{vis}(i)})^2 + m_{\PSGc}^2
	+ 2 \left( E_{\mathrm{T}}^{\text{vis}(i)} E_{\mathrm{T}}^{\PSGc{(i)}}
	- \vec{p}_{\mathrm{T}}^{\; \text{vis}(i)} \cdot \vec{p}_{\mathrm{T}}^{\; \PSGc{(i)}}
     \right).
\label{eq.MT2.transmass}
\end{eqnarray}
These have the property (as in W-boson decays)
that, for the true LSP mass $m_{\PSGc}$, their distribution cannot exceed the mass of the parent particle of the decay
and they present an endpoint at the value of the parent mass.
The momenta $\vec{p}_{\mathrm{T}}^{\; \PSGc(i)}$ of the invisible particles are not experimentally accessible individually.
Only their sum, the missing transverse momentum $\vec{p}_{\mathrm{T}}^{\; \text{miss}}$, is known.
Therefore, in the context of SUSY, a generalization of the transverse mass is needed
and the proposed variable is \MTtwo.
It is defined as
\begin{eqnarray}
	\MTtwo(m_{\PSGc}) = \min_{\vec{p}_{\mathrm{T}}^{\; \PSGc(1)} + \vec{p}_{\mathrm{T}}^{\; \PSGc(2)} = \vec{p}_{\mathrm{T}}^{\; \text{miss}}}
	\left[ \max \left( M_{\mathrm{T}}^{(1)} , M_{\mathrm{T}}^{(2)} \right) \right] ,
\label{eq.MT2.definition}
\end{eqnarray}
where the LSP mass $m_{\PSGc}$ remains a free parameter.
This formula can be understood as follows.
As neither $M_{\mathrm{T}}^{(1)}$ nor $M_{\mathrm{T}}^{(2)}$ can exceed the parent mass if the true momenta are used,
the larger of the two can be chosen.
To make sure that \MTtwo does not exceed the parent mass,
a minimization is performed on trial LSP momenta fulfilling the \ptmissvec constraint.
The distribution of \MTtwo for the correct value of $m_{\PSGc}$
then has an endpoint at the value of the primary particle mass.
If, however,  $m_{\PSGc}$ is lower (higher) than the correct mass value, the endpoint will be below (above) the parent mass.
An analytic expression for \MTtwo has been computed \cite{Cho:2007dh} assuming that
initial-state radiation (ISR) can be neglected.
In practice, the determination of \MTtwo may be complicated by
the presence of ISR
or, equivalently, transverse momentum arising from decays that occur upstream
in the decay chain \cite{Burns:2008va}.
In this case, no analytic expression for \MTtwo is known,
but it can be computed numerically, using, e.g., the results of Ref.~\cite{Cheng:2008hk}.

To illustrate the behavior of \MTtwo,
we consider the simple example of \MTtwo without ISR or upstream transverse momentum.
As discussed in Ref.~\cite{Cho:2007dh}, the angular and \pt dependence
of \MTtwo is encoded in a variable $A_{\mathrm{T}}$:
\begin{eqnarray}
	A_{\mathrm{T}} &=& E_{\mathrm{T}}^{\mathrm{vis}(1)} E_{\mathrm{T}}^{\mathrm{vis}(2)} + \vec{p}_{\mathrm{T}}^{\; \mathrm{vis}(1)} \cdot \vec{p}_{\mathrm{T}}^{\; \mathrm{vis}(2)},
\label{eq.MT2.AT}
\end{eqnarray}
and \MTtwo increases as $A_{\mathrm{T}}$ increases.
Therefore, the minimum value of \MTtwo is reached in configurations
where the visible systems are back-to-back.
The maximum value is
reached when they are parallel to each other and have large \pt.
In the simple case where $m_{\PSGc} = 0$ and the visible systems have zero mass,
\MTtwo becomes
\begin{eqnarray}
        \label{eq:MT2simplified}
	(\MTtwo)^2 = 2 A_{\mathrm{T}} = 2  p_{\mathrm{T}}^{\mathrm{vis}(1)} p_{\mathrm{T}}^{\mathrm{vis}(2)} ( 1 + \cos \phi_{12} ) ,
\end{eqnarray}
where $\phi_{12}$ is the angle between the two visible systems in the transverse plane.
It can be seen that Eq.~\eqref{eq:MT2simplified} corresponds to the transverse mass of two systems
$(\MT)^2 = 2 p_{\mathrm{T}}^{\mathrm{sys}(1)} p_{\mathrm{T}}^{\mathrm{sys}(2)} (1 - \cos \phi_{12})$,
with $\vec{p}_{\mathrm{T}}^{\, \text{vis}}=-\vec{p}_{\mathrm{T}}^{\, \text{sys}}$ for one of the systems.

In this paper, we use \MTtwo as a variable
to distinguish potential new physics events from
SM backgrounds.
The use of \MTtwo as a discovery variable was first proposed in Ref.~\cite{Barr:2009wu} , but
here we follow a different approach.
Several choices for the visible system used as input to \MTtwo can be considered:
dijet events
(as in Ref.~\cite{Barr:2009wu}),
the two jets with largest \pt in multijet events,
or two systems of pseudojets defined by grouping jets together.
In this study, we use the last method.

A technique to group jets in multijet events into two pseudojets
is the ``event hemispheres'' method described in Ref.~\cite{CMSTDR07} (see Section 13.4).
We take the two initial axes as the directions of the two massless jets that yield the largest dijet invariant mass.
The pseudojets are then formed based on a minimization of the Lund distance criterion \cite{CMSTDR07,Sjostrand:2006za}.

We use \MTtwo as
our main search variable since
SUSY events
with large expected \MET and jet acoplanarity
will be concentrated in the large \MTtwo region.
In contrast, QCD dijet events, in which the two jets are back-to-back, populate the region of small \MTtwo
regardless of the value of \MET or jet \pt.
In the present study, we choose the visible systems to be massless and
set $m_{\PSGc} = 0$.
Then back-to-back dijet events will have $\MTtwo=0$, as explained above.
Hence, \MTtwo has a built-in
protection against jet mismeasurements in QCD dijet events,
even if accompanied by large \MET.
However, QCD multijet events with large \MET may give rise to acoplanar pseudojets, leading to larger \MTtwo values.
For this reason, further protections against \MET from
mismeasurements need to be introduced, as described below.
Other SM backgrounds, such as \ttbar, single top-quark, and W+jets events with
leptonic decays, or Z+jets events where the Z boson decays to neutrinos, contain
true \MET and can also lead to acoplanar pseudojets.

\section{CMS detector}
\label{sec:CMS}

The central feature of the CMS apparatus is a superconducting solenoid 13\unit{m} in length and 6\unit{m} in diameter
that provides an axial magnetic field of 3.8\unit{T}.
The core of the solenoid is instrumented with various particle detection systems: a silicon pixel
and strip tracker, an electromagnetic calorimeter (ECAL), and a brass/scintillator hadron calorimeter
(HCAL).  The silicon pixel and strip tracker covers $|\eta|<2.5$,
where pseudorapidity $\eta$ is defined by $\eta = -\ln{[\tan{(\theta/2)}]}$ with $\theta$ the polar angle
of the trajectory of the particle with respect to the counterclockwise beam direction.
The ECAL and HCAL cover $|\eta|<3$. The steel return yoke outside
the solenoid is instrumented with gas detectors used to identify muons.
A quartz-steel Cerenkov-radiation-based forward hadron calorimeter extends the coverage to $|\eta| \leq 5$.
The detector is nearly hermetic, covering $0<\phi<2\pi$ in azimuth,
allowing for energy balance measurements in the plane transverse to the beam
directions.
The first level of the CMS trigger system, composed of custom hardware
processors, uses information from the calorimeters and muon detectors to
select the most interesting events in a fixed time interval of less than
4\mus. The High Level Trigger processor farm further decreases the event
rate from around 100\unit{kHz} to around 300\unit{Hz}, before data storage.
A detailed description of the CMS detector can be found elsewhere \cite{CMSdet}.

\section{Samples and event selection}
\label{sec:cut-flow}
The data used in this analysis were collected by triggers based on the
quantity \HT, the scalar sum of transverse momenta of reconstructed and
energy-corrected calorimeter jets.
Due to a continuous increase in the instantaneous luminosity of the LHC,
the trigger evolved with time from the requirement $\HT > 440$\GeV to $\HT >
750$\GeV.
In this analysis, only triggers
with a threshold of 650\GeV or less have been used, corresponding to a
total integrated luminosity of \Lumino.

The analysis is designed using simulated event samples created with
the \PYTHIA6.4.22  \cite{Sjostrand:2006za} and \MADGRAPH5v1.1  \cite{Alwall:2011uj}
Monte Carlo %(MC) not needed anymore
event generators. These events are subsequently processed
with a detailed simulation of the CMS detector response based on \GEANTfour \cite{Geant4}.
The events are reconstructed and analyzed in the same way as the data.
The SUSY signal particle spectrum is calculated using
{\textsc{softsusy}\xspace}~\cite{Allanach2002305} and
for the decays {\textsc{sdecay}\xspace}~\cite{Muhlleitner:2003vg}  is used.
We use two CMS SUSY benchmark signal samples, referred to as LM6 and LM9 \cite{CMSTDR07}, to illustrate
possible CMSSM~\cite{CMSSM} yields.
The CMSSM is defined by the universal scalar and gaugino mass parameters $m_0$ and $m_{1/2}$, respectively,
the parameter $A_0$ of the trilinear couplings, the ratio  of the vacuum expectation values
of the two Higgs fields $\tan{\beta}$, and the sign of the Higgs mixing parameter $\sign{(\mu)}$.
The parameter values for LM6 are $m_0 = 85$\GeV, $m_{1/2} = 400$\GeV, $\tan\beta = 10$,
$A_0=0$\GeV  and $\sign( \mu) > 0$. Those for LM9 are $m_0 = 1450$\GeV, $m_{1/2} = 175$\GeV, $\tan\beta = 50$,
$A_0=0$\GeV  and $\sign(\mu) > 0$.
All samples are generated using the CTEQ6~\cite{CTEQ6} parton distribution functions (PDFs).
For SM background  simulated samples we use  the most accurate calculation
of the cross sections currently available, usually with next-to-leading-order (NLO) accuracy.
For the CMS SUSY benchmark signal samples
we use NLO cross sections of 0.403\unit{pb} and 10.6\unit{pb} for LM6 and LM9, respectively,
obtained by weighting the leading order cross sections from \PYTHIA with sub-process dependent K-factors calculated with \PROSPINO~\cite{Prospino97}.

The events are reconstructed
using the particle-flow (PF) algorithm~\cite{pflow09},
which identifies and reconstructs individually the particles produced in the collision,
namely charged hadrons, photons, neutral hadrons, electrons, and muons.

Electrons and muons with $\pt \geq 10$\GeV and $|\eta| \leq 2.4$ are
considered isolated if
the transverse momentum sum of charged hadrons, photons, and neutral hadrons surrounding the lepton
within a cone of radius $\sqrt{(\Delta \eta)^2 + (\Delta \phi)^2}=0.4$,
divided by the lepton transverse momentum value itself, is less than 0.2.
The electron and muon reconstruction and identification algorithms are
described in Refs.~\cite{EGM-10-004,PFT-10-003} and~\cite{MUO-10-002},
respectively.
All particles apart from the isolated electrons and muons are clustered into jets using the anti-$k_\mathrm{T}$ jet
clustering algorithm~\cite{Cacciari:2008gp} with distance parameter $0.5$ \cite{JME-10-003, PFT-10-002}.
Jet energies are calibrated by applying correction factors as a function of the transverse momentum and the pseudorapidity of the jet.
Residual jet energy corrections are applied to jets in data to account for differences in jet energy scale between simulation and data \cite{CMS-JetEnergy}.
The effect of pileup, namely multiple pp collisions within a beam crossing, is reduced
by using the FastJet pileup subtraction procedure
\cite{Cacciari:2007fd,Cacciari:2008gn} for data and simulated events.
Jets are required to pass loose identification criteria and to satisfy $\pt>20$\GeV and $|\eta| \leq 2.4$.
The b-jet tagging is based on the simple-secondary-vertex
algorithm~\cite{PAS-BTV-10-001}. We use the high-purity working point that yields a typical jet-tagging efficiency of 42\% for b jets in our search region
while the mistagging efficiency for light-flavored (uds quark and gluon) jets
is of the order of 0.1\% and for c jets, 6.3\%.
The missing transverse momentum \METvec is computed as the negative vector sum of all particles reconstructed by the PF algorithm~\cite{PFT-10-002}.

Events are required to contain at least one good primary vertex \cite{CMS-PAS-TRK-10-005}.
The \HT value, computed from PF jets with $\pt > 50$\GeV, must satisfy $\HT \geq 750$\GeV.
With this \HT requirement, the triggers are nearly 100\% efficient.
At least three jets are required, where a \pt threshold of 40\GeV is used for jet counting.
The two leading jets are required to have $\pt > 100$\GeV.
The value of \MET is required to exceed 30\GeV.
Events containing beam background or anomalous calorimeter noise are rejected.
To reject events where a significant fraction of the momentum imbalance
arises from forward or soft jets, a maximum difference of 70\GeV is imposed
on the modulus of the difference between the \METvec and \MHTvec vectors,
where \MHTvec is the negative vector sum of all selected jets.
Events containing jet candidates with $\pt>50$\GeV that fail the
jet identification criteria are also rejected.

To reduce the background from QCD multijet events with large \MET,
arising from mismeasurements or leptonic heavy flavor decays,
a minimum azimuthal difference $\Delta\phi_{\mathrm{min}}{(\mathrm{jets}, \METvec)}>0.3$
is required between
the directions of \METvec and any jet with $\pt > 20\GeV$.
Finally, events are rejected if they contain an isolated electron or muon,
to suppress the contributions from W+jets, Z+jets and top-quark backgrounds.

\section{Search strategy}
\label{sec:strategy}

The \MTtwo variable
is computed after applying the selection criteria of Section~\ref{sec:cut-flow}.
We separately consider fully hadronic channels with ${\geq}3$ jets and
a tight \MTtwo requirement (the \MTtwo analysis), which is mostly sensitive to signal
regions with large squark and gluino masses, and channels with ${\geq}4$ jets, at least one
tagged b jet, and a relaxed \MTtwo requirement (the \MTtwob analysis), which increases
sensitivity to regions with small gluino and large squark masses.

Given the event selection outlined above, we do not expect a significant number of QCD multijet events to appear in the signal regions.
Nonetheless, we estimate an upper limit
on the remaining QCD multijet background in the signal regions from data control samples.
The main backgrounds, consisting of W+jets, Z+jets, and top-quark production,
are evaluated from data control samples and simulation.
A common strategy is applied to both the \MTtwo  and \MTtwob analyses:
\begin{itemize}
\item 	Two regions are defined in \HT, a low \HT region $750 \leq \HT < 950$\GeV
	and a high \HT region $\HT\geq 950\GeV$.
	In each region, several adjacent bins in \MTtwo are defined:
	five bins for the \MTtwo analysis and four for the \MTtwob analysis.
	The lowest bin in \MTtwo is chosen such that
	the expected QCD multijet background remains a small fraction of the total background.
	For the \MTtwo analysis the lowest bin starts at $\MTtwo = 150$\GeV and for \MTtwob at $\MTtwo = 125$\GeV.

\item 	A dedicated method for each background is designed to estimate its
	contribution in the signal region from data control samples and simulation.
	The number of
	events and their relative systematic uncertainties
	are computed by means of these methods in each $\HT$, $\MTtwo$ bin.
	The methods are designed such that the resulting estimates are largely uncorrelated statistically.

\item	The predicted number of events for all background components and their uncertainties
	are combined, resulting in an estimate of the total background yield and its
	uncertainty in each bin.
\item 	The estimated number of background events for each bin is compared to
	the number of observed events, and the potential contribution from a SUSY
	signal is quantified by a statistical method described in Section~\ref{sec:exclusion}.
\end{itemize}

\section{\texorpdfstring{\MTtwo}{MT2} analysis}
\label{sec:data:geq3jets}

\begin{figure}[!htb]
	\begin{center}
	\includegraphics[width=0.7\textwidth]{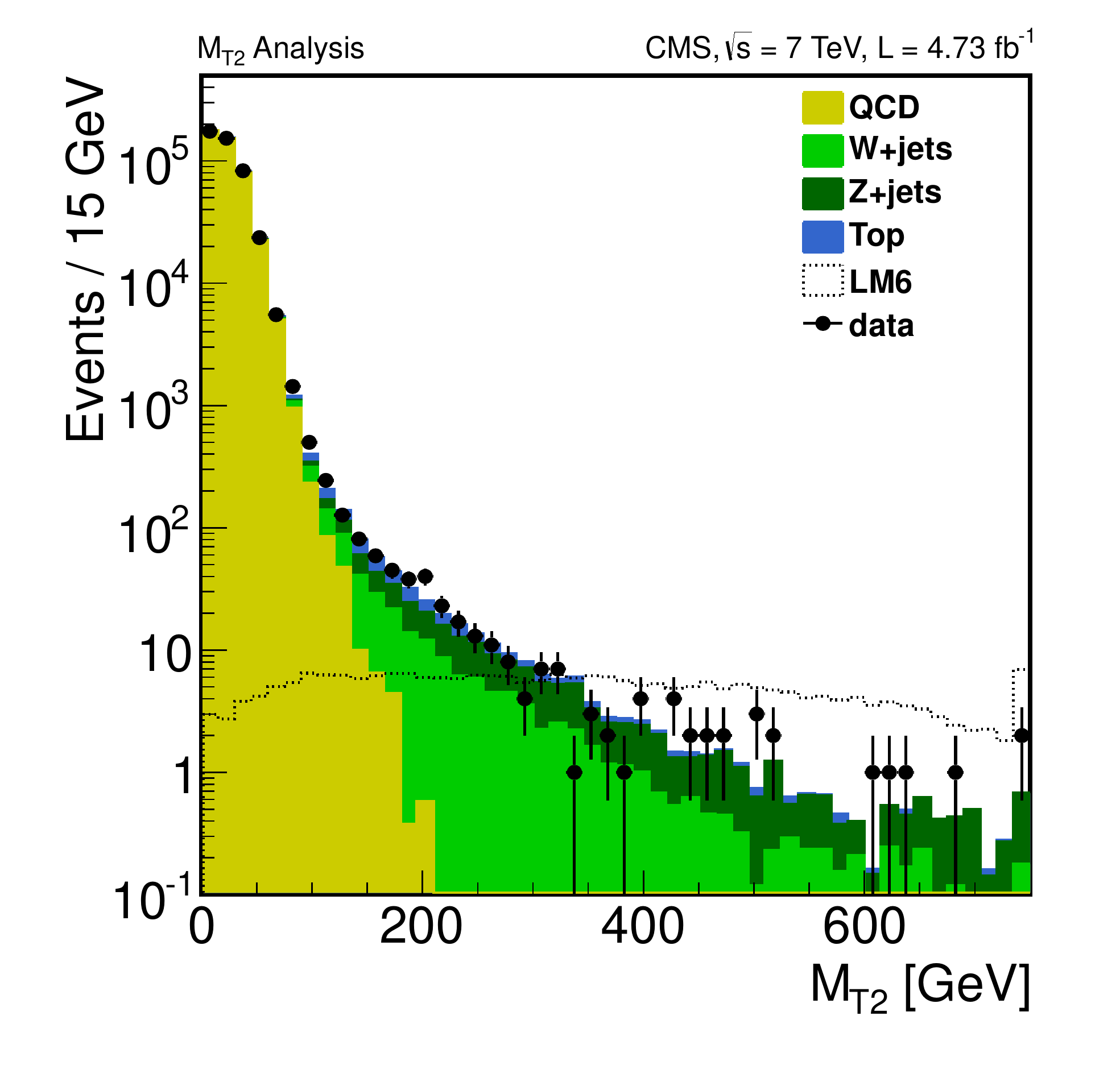}
	\caption{The \MTtwo distribution with all selection requirements applied and $\HT \geq 750\GeV$.
	The different predictions for the SM backgrounds from simulation are stacked on top
	of each other.  The LM6 signal distribution is not stacked.  All
	distributions from simulation are normalized to the integrated luminosity of the data.}
	\label{fig:MT2_combinedHT}
    \end{center}
\end{figure}
\begin{table}[h!]
\begin{center}
\topcaption{Observed number of events and expected SM background yields from simulation in	
\MTtwo bins for the low and high \HT regions. These numbers are for guidance only and are not used in the final background
prediction.}
\label{table:cut-flow-SMdata}
\begin{tabular}{lcccc|cccc}
\hline\hline
& QCD multijet& $W+$jets& Top & $Z(\nu\nu)+$jets& Total SM & Data \\ \hline \hline
 $750 \leq \HT < 950$ \\ \hline
 $\MTtwo [0,\infty]$ & 3.18e+05 & 9.22e+02 & 1.30e+03 & 3.01e+02 & 3.20e+05 & 3.20e+05 \\
 $\MTtwo [150, 200]$ & 3.08 & 37.5 & 20.6 & 27.9 & 90.0 & 88 \\
 $\MTtwo [200, 275]$ & 0.0 & 20.6 & 9.40 & 20.3 & 50.3 & 69 \\
 $\MTtwo [275, 375]$ & 0.0 & 9.74 & 2.74 & 11.6 & 24.1 & 19 \\
 $\MTtwo [375, 500]$ & 0.0 & 3.63 & 0.69 & 6.07 & 10.4 & 8 \\
 $\MTtwo [500, \infty]$ & 0.0 & 1.54 & 0.20 & 3.55 & 5.29 & 6 \\
\hline
         $\HT \geq 950$ \\ \hline
\hline
 $\MTtwo [0,\infty]$ & 1.22e+05 & 4.39e+02 & 6.32e+02 & 1.42e+02 &  1.23e+05 & 1.19e+05 \\
 $\MTtwo [150, 200]$ & 9.84 & 19.8 & 11.7 & 12.9 &  54.2 & 70 \\
 $\MTtwo [200, 275]$ & 0.47 & 13.7 & 5.25 & 10.5 &  30.0 & 23 \\
 $\MTtwo [275, 375]$ & 0.04 & 6.43 & 1.83 & 6.42 &  14.7 & 9 \\
 $\MTtwo [375, 500]$ & 0.0  & 1.63 & 0.40 & 2.54 &  4.57 & 8 \\
 $\MTtwo [500, \infty]$ & 0.0 & 1.10 & 0.16 & 2.16 &  3.42 & 4 \\
\hline\hline
\end{tabular}
\end{center}
\end{table}

Figure \ref{fig:MT2_combinedHT}
shows the measured \MTtwo distribution in comparison to simulation.
For $\MTtwo<80$\GeV the distribution is completely dominated by
QCD multijet events.  For medium \MTtwo values, the
distribution is dominated by W+jets and Z$(\cPgn\cPagn)$+jets events
with some contribution from top-quark events,
while in the tail of \MTtwo the contribution from top-quark
production becomes negligible and Z$(\cPgn\cPagn)$+jets together with W+jets
events dominate.
We observe good agreement between data and simulation in the core as well as in the tail of the distribution.
The white histogram (black dotted line) corresponds to the LM6 signal.
It can be noted that in the presence of signal, an excess in the tail of \MTtwo
is expected.

The corresponding event yields for data and SM simulated samples,
after the full selection and
for the various bins in \MTtwo, are given in Table~\ref{table:cut-flow-SMdata}
for the low and the high \HT regions.
Contributions from other backgrounds, such as $\gamma$+jets, $Z(\ell\ell)$+jets and
diboson production, are found to be negligible.
It is seen that for all but one \MTtwo bin,
the observed number of events agrees within the uncertainties with the SM
background expectation from simulation.
In the low \HT region, the \MTtwo bin $[200, 275]\GeV$ exhibits an excess in data compared to background.
We investigated whether the origin could be instrumental in nature,
but did not find evidence for it.
It could be of statistical origin.
The excess has a marginal impact on the final observed limit.

\subsection{Background prediction}
\label{sec:data:geq3jets:bkgd}

\subsubsection{QCD multijet background}

The simulation predicts that the QCD multijet background is negligible in the tail
of the \MTtwo distribution.
Nevertheless, a dedicated method using a data control region was designed to
verify that this is indeed the case.

We base this estimation on \MTtwo and $\Delta \phi_{\mathrm{min}}$, which is
the difference in azimuth between \METvec and the closest jet.
The background in the signal region, defined by $\Delta\phi_{\mathrm{min}} \geq 0.3$ and large \MTtwo,
is predicted from a control region with $\Delta\phi_{\mathrm{min}} \leq 0.2$.
The two variables are strongly correlated, but a factorization method can still be applied if the functional form is known for the ratio of the number of events
$r(\MTtwo) = N(\Delta\phi_{\mathrm{min}} \geq 0.3) / N(\Delta\phi_{\mathrm{min}} \leq 0.2)$
as a function of \MTtwo.
It is found from simulation studies, and confirmed with data, that for $\MTtwo > 50$\GeV the ratio falls exponentially.
Therefore, a parameterization of the form
\begin{eqnarray}
	r(\MTtwo) = \frac{N(\Delta\phi_{\mathrm{min}} \geq 0.3)}{N(\Delta\phi_{\mathrm{min}} \leq 0.2)}
	= \exp{(a - b \MTtwo)} + c
\label{eq.bkgdpred.ratio}
\end{eqnarray}
is used for $\MTtwo > 50$\GeV.
The function is assumed to reach a constant value at large \MTtwo due to
extreme tails of the jet energy resolution response.

The method is validated with simulation. First the parameters $a$,
$b$, and $c$ are extracted from a fit to simulated QCD multijet events in the full
$\MTtwo$  spectrum.
The fitted parameter value for $c$ is compatible with a negligible
QCD multijet contribution at large \MTtwo.
It is verified that similar fit results for the parameters $a$ and $b$
are obtained when the fit is limited to the region $50 < \MTtwo < 80$\GeV,
where contributions from background processes other than that from QCD
multijets is small. The robustness of the prediction is checked by
systematically varying the fit boundaries.

For the final results, we repeat the fit to data in the region
$50<\MTtwo<80$\GeV, after subtracting the W+jets, Z+jets and top background
contributions using simulation. The fitted parameter values for $a$ and $b$ are
in agreement with the values obtained from the QCD multijet simulation.
We conservatively fix the constant $c$ to the value of the exponential
at $\MTtwo = 250$\GeV, where agreement with data can still be verified.
In the lower \MTtwo bins, where the exponential term dominates,
the method reliably predicts the QCD multijet background.
For higher \MTtwo bins, where the constant term dominates, the method
overestimates the number of QCD multijet events relative to the simulation, nonetheless
confirming that the QCD multijet contribution is negligible.

The extreme case of total loss of a jet, leading to population of the high \MTtwo tail,
is studied using a sample of high \pt mono-jet events obtained with a dedicated event selection.
The total number of events is found to be compatible within the uncertainties
with the number expected from the electroweak processes,
confirming that the QCD multijet contribution is negligible and hence that the constant $c$ is small.

\subsubsection{ W$(\ell\nu)$+jets and top-quark background}

The backgrounds due to W$(\ell\nu)$+jets and to semi-leptonic decays of top quarks have the following sources in common:
\begin{itemize}
	\item 	leptonic decays of the W boson, where the lepton is unobserved because it falls outside the \pt or $\eta$ acceptance;
	\item 	to a lesser extent, leptonic decays of the W boson, where the lepton is within the acceptance
		but fails to satisfy the reconstruction, identification, or isolation criteria;
	\item 	W$(\tau \nu_{\tau})$ decays, where the $\tau$ decays hadronically.
\end{itemize}
We refer to leptons that fall into either of the first two categories as ``lost leptons''.
The number of events with lost leptons is estimated from a data control
sample where a single lepton (e or $\mu$) is found.
A correction factor accounting for the probability to lose the lepton is
derived from simulation.
To avoid a potential contamination from signal events,
a transverse mass cut $\MT < 100$\GeV is introduced.
This method is applied in the various \HT and \MTtwo bins.
First, a successful validation test of the method is performed using simulated
samples.
Then, a prediction is made from the data bin by bin and found to be in
agreement with the expectation from simulation.
A systematic uncertainty is evaluated that includes the uncertainty on the lepton efficiencies, acceptance, and background subtraction.

For the background contribution from hadronically decaying tau leptons, a method similar to the one described above is used.
Events with an isolated and identified hadronically decaying
tau~\cite{PFT-11-001} lepton
are selected in the various \HT and \MTtwo bins. The contribution
from jets misidentified as taus is subtracted.
The remaining number of tau events is corrected for the tau reconstruction and identification efficiency.
The predicted number of hadronically decaying tau background events agrees with the true number from simulation. Given the small
number of events in the data,
the numbers of events from the simulation are used for the background estimate,
with the same relative systematic uncertainties as for the lost leptons.

\subsubsection{Z$(\nu \bar{\nu})$+jets background}

The estimate of the Z$(\cPgn\cPagn)$+jets background is obtained
independently from two distinct data samples, one containing $\gamma$+jets
events and the other W$(\mu \nu)$+jets events. In both cases the
invisible decay of the Z boson is mimicked by removing, respectively, the
photon and the muon from the event, and adding vectorially the corresponding \ptvec to
\METvec.

For the estimate based on $\gamma$+jets events, a sample of events
with identified and isolated photons~\cite{EGM-10-006} with $\pt > 20$\GeV is selected,
where all selection requirements except that on \MTtwo are imposed.
This sample contains both prompt photons and photons from $\pi^0$ decays in QCD multijet events.
The two components are separated by performing a maximum likelihood fit of
templates from simulated events
to the shower shapes. The event sample is dominated by
low \pt photons, where the shower shape provides high discrimination
power between prompt photons and $\pi^0$s. The extrapolation of their contributions
as a function of \MTtwo is obtained from simulation.
The Z$(\cPgn\cPagn)$+jets background is estimated for each bin in \MTtwo from the number
of prompt photon events
multiplied by the \MTtwo-dependent ratio of Z$(\cPgn\cPagn)$+jets to
$\gamma$+jets events  obtained from simulation.
This ratio increases as a function of the photon \pt
(which drives the \MTtwo value)
and reaches a constant value above 300\GeV.
The resulting prediction of the background is found to be in good
agreement with the expectation from simulation.
Systematic uncertainties on the background prediction consist of
the statistical uncertainties from the number of $\gamma$+jets events,
a normalization uncertainty in the shower shape fit of~5\%,
and the systematic uncertainties on the ratio of Z$(\cPgn\cPagn)$+jets to
$\gamma$+jets events in the simulation.
The uncertainties on the ratio are estimated to be less than 20\% (30\%) for $\MTtwo < 275$ ($\MTtwo > 275$)\GeV.
To assess these uncertainties, the \pt dependence of the ratio is studied in data and compared to simulation using
leptonically decaying Z events.
For $\pt >  400$\GeV this test is limited by the number of the leptonic $Z$ events,
which justifies the increased uncertainty for $\MTtwo > 275$\GeV.

For the estimate from W$(\mu \nu)$+jets events,
corrections are needed for lepton acceptance, lepton reconstruction efficiency,
and the ratio between the
production cross sections for W and Z bosons
(including differences between the shapes of the distributions on which selection criteria are applied).
The lepton efficiencies are taken from studies of Z$(\mu\mu)$ events in data.
Also, the top-quark background to the W+jets sample is subtracted.
The top-quark background is evaluated by applying b
tagging to the data to identify top-quark decays and then correcting for the b-tagging
efficiency.
The Z$(\cPgn\cPagn)$+jets background is then estimated in each of the \MTtwo bins.
The systematic uncertainty includes the contributions from the lepton
selection and reconstruction efficiencies, the the b-tagging efficiency, the
acceptance from simulation, and the W-to-Z ratio.

The Z$(\cPgn\cPagn)$+jets background estimates
from the $\gamma$+jets and W$(\mu \nu)$+jets
methods are in good agreement with each other.
Since they are statistically uncorrelated, we take the weighted
average of the two predictions as the final estimate.

\subsection{Results}
\label{sec:data:geq3jets.results}

\begin{figure}[!htb]
  \centering
  \includegraphics[width=0.49\textwidth]{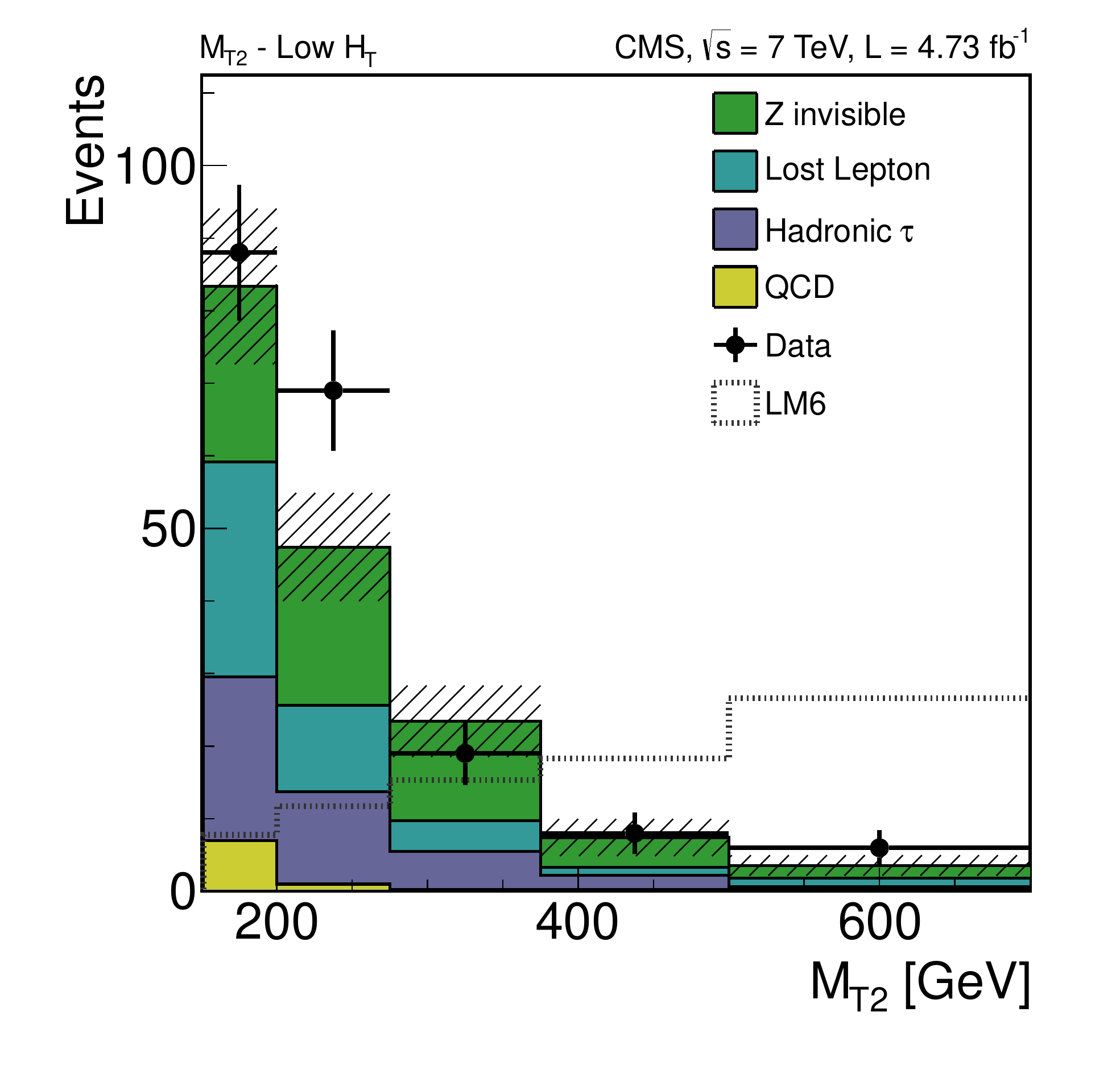}
  \includegraphics[width=0.49\textwidth]{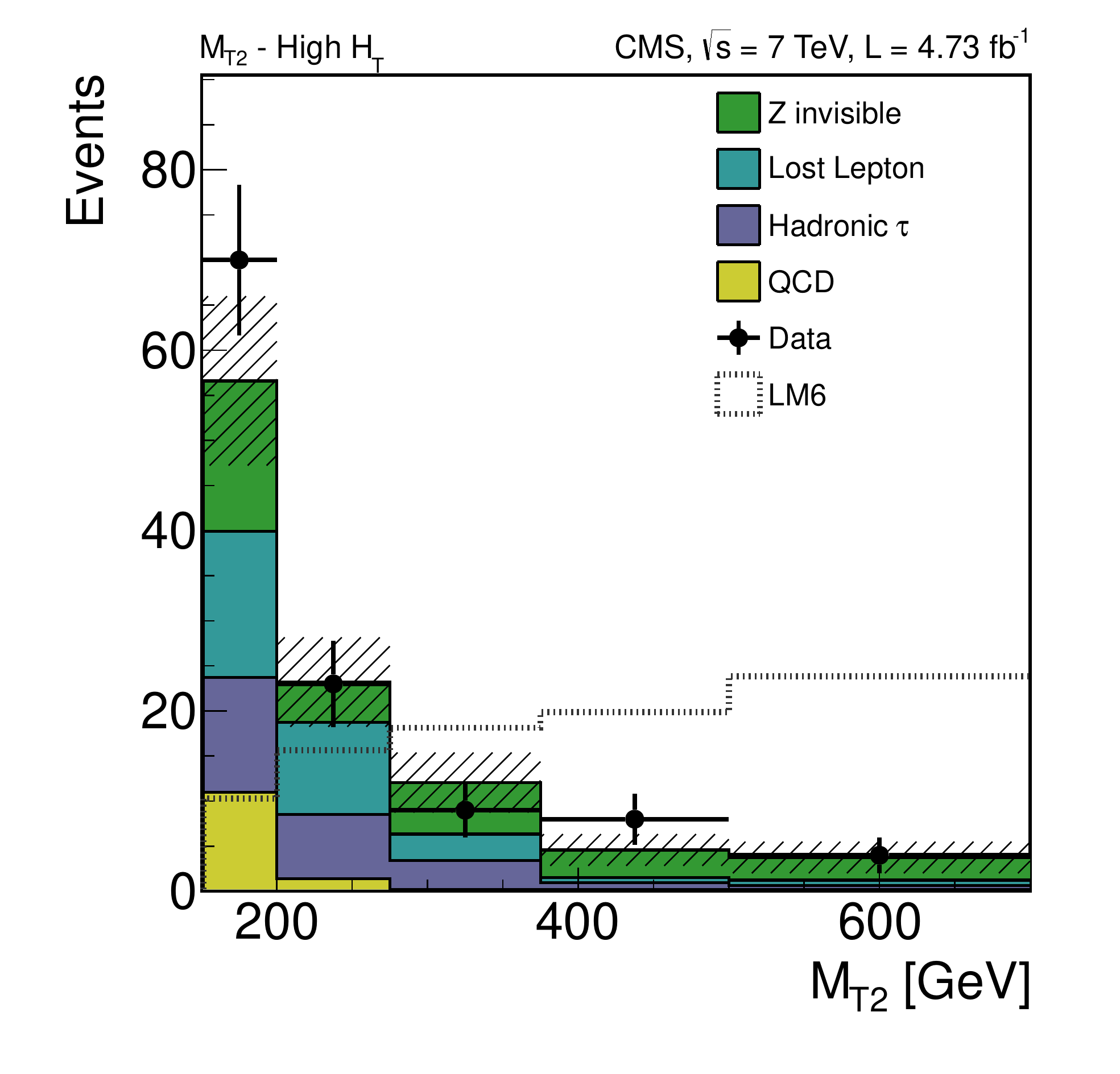}
  \caption{\MTtwo distribution from the background estimates
    compared to data. The figure on the left corresponds to the
    $750\leq \HT<950$\GeV region, while that on the right corresponds to $\HT \geq 950\GeV$.
    The predictions from simulated events for the LM6 signal model
     (not stacked) are also shown.
The hatched band shows the total uncertainty on the SM background estimate.}
  \label{fig:datadrivenMT2}
\end{figure}

The results of the background estimation methods for each background
contribution are summarized in Table~\ref{table:datadrivenMT2}
and shown in Fig.~\ref{fig:datadrivenMT2}.
\begin{table}[h!]
  \begin{footnotesize}
    \begin{center}
      \topcaption{Estimated event yields for each background contribution in
        the various \MTtwo and \HT bins. The predictions from control regions
        in data are compared to the expected event yields from simulation.
        Statistical and systematic uncertainties are added in quadrature. The
        total background prediction is compared to data in the last two
        columns.}
	\label{table:datadrivenMT2}
	\begin{tabular}{@{ }l|cc|cc|c|cc|c|c@{ }}
	\hline\hline
	& \multicolumn{2}{c|}{$Z\to\nu\bar{\nu}$} & \multicolumn{2}{c|}{Lost lepton} & $\tau\to \mathrm{had}$ & \multicolumn{2}{c|}{QCD multijet} & Total bkg. & Data \\
	            & sim. & data pred. & sim. & data pred. & Estimate & sim. & data pred. &  data pred. & \\ \hline
	$750 \leq \HT < 950$ \\ \hline
	$\MTtwo[150,200]$   & 27.9  & $24.2\pm4.9$ & 36.0 & $29.6\pm7.1$ & $22.5\pm5.4$ & 3.1 & $7.0\pm3.5$    & $83.3\pm10.7$ & 88\\
	$\MTtwo[200,275]$   & 20.3  & $21.8\pm4.8$ & 17.2 & $11.9\pm3.9$ & $12.7\pm4.2$ & 0.0 & $1.0\pm0.5$    & $47.4\pm7.5$  & 69\\
	$\MTtwo[275,375]$   & 11.6  & $13.7\pm3.8$ & 7.1  & $4.2\pm1.9$  & $5.4\pm2.5$  & 0.0 & $0.14\pm0.07$  & $23.4\pm4.9$  & 19\\
	$\MTtwo[375,500]$   & 6.1   & $4.1\pm1.6$  & 2.2  & $1.1\pm0.9$  & $2.2\pm1.8$  & 0.0 & $0.08\pm0.05$  & $7.4\pm2.6$   & 8\\
	$\MTtwo[500,\infty]$& 3.5   & $1.8\pm0.9$  & 1.1  & $1.2\pm1.0$  & $0.6\pm0.5$  & 0.0 & $0.00\pm0.00$  & $3.6\pm1.4$   & 6\\
	\hline
	$\HT \geq 950$ \\ \hline
	$\MTtwo[150,200]$  & 12.9 & $16.7\pm3.6$ & 18.7 & $16.2\pm5.3$ & $12.7\pm4.1$ & 9.8  & $11.0\pm5.5$   & $56.6\pm9.4$ & 70\\
	$\MTtwo[200,275]$  & 10.5 & $4.5\pm2.0$  & 11.7 & $10.2\pm3.7$ & $7.1\pm2.6$  & 0.47 & $1.4\pm0.7$    & $23.2\pm5.0$ & 23\\
	$\MTtwo[275,375]$  & 6.4  & $5.7\pm2.2$  & 5.0  & $2.9\pm1.7$  & $3.3\pm1.9$  & 0.04 & $0.13\pm0.07$  & $12.1\pm3.3$ & 9\\
	$\MTtwo[375,500]$  & 2.5  & $3.0\pm1.4$  & 1.1  & $0.6\pm0.6$  & $0.9\pm0.9$  & 0.0  & $0.06\pm0.04$  & $4.6\pm1.8$  & 8\\
	$\MTtwo[500,\infty]$& 2.2  & $2.5\pm1.5$  & 0.6  & $0.6\pm0.6$  & $0.6\pm0.6$  & 0.0  & $0.06\pm0.04$  & $3.8\pm1.7$  & 4\\
	\hline\hline
	\end{tabular}
	\end{center}
	\end{footnotesize}
\end{table}

\section{\texorpdfstring{\MTtwob}{MT2b} analysis}
\label{sec:data:relaxed}

The selection criteria developed for the
\MTtwo analysis are not optimal for events with heavy squarks and light gluinos,
such as are predicted by the SUSY benchmark model LM9.
To improve sensitivity to these types of events, we perform the \MTtwob analysis
based on loosened kinematic selection criteria and the requirement of a tagged
b jet.
The restriction on \MTtwo is loosened to $\MTtwo>125\GeV$ and the $\Delta\phi_{\text{min}}{(\text{jets}, \METvec)}>0.3$
requirement is applied to the four leading jets only.
We require that there be at least four jets with $\pt >40\GeV$,
and the leading jet to have $\pt > 150\GeV$.
We further require that at least one of the jets in the event be tagged as a b-quark jet.

\begin{figure}[!htb]
\begin{center}
 \includegraphics[width=0.7\textwidth]{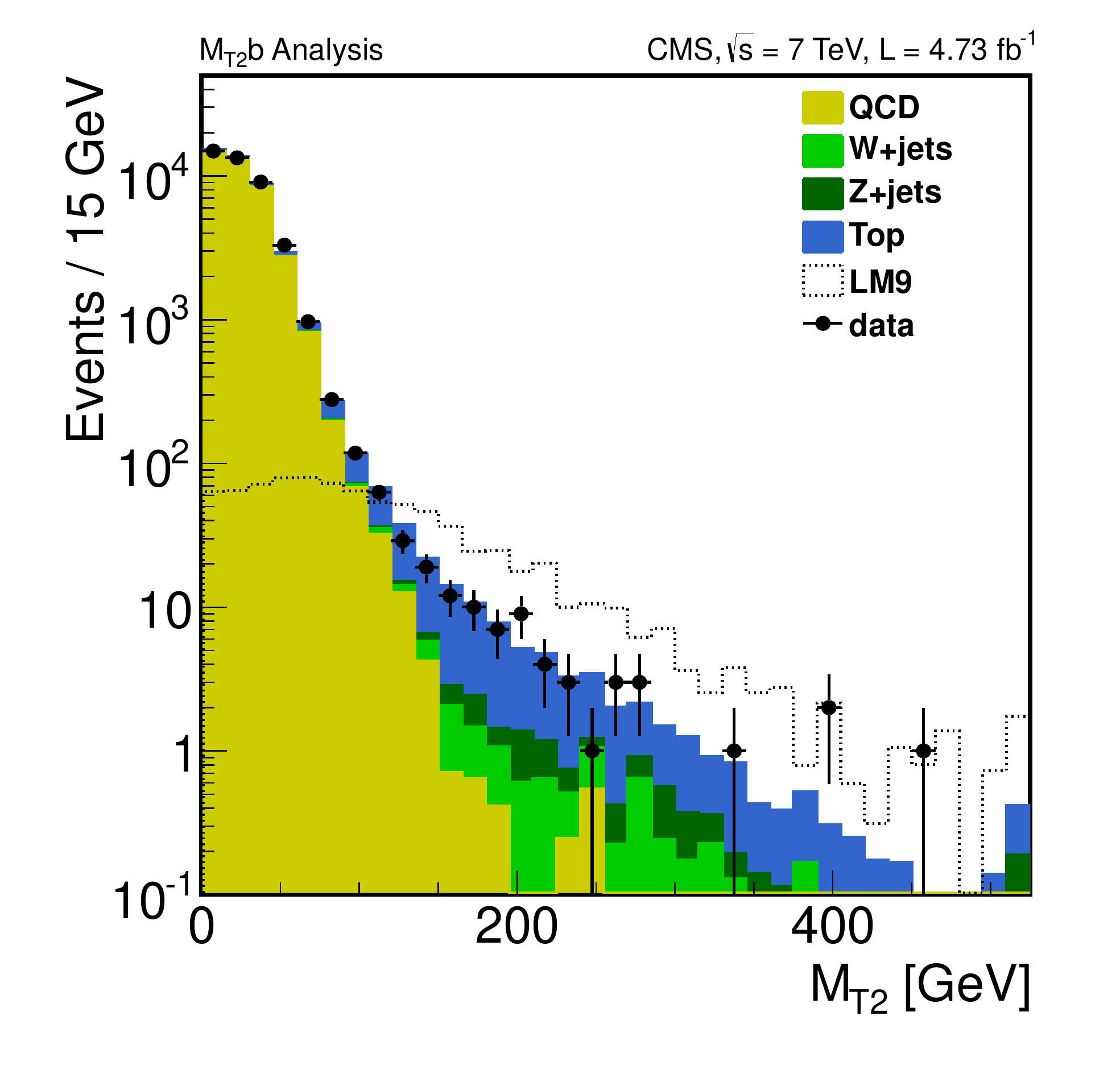}
\caption{
\MTtwo for events with the \MTtwob selection criteria applied
and with $\HT \geq 750\GeV$.
The different predictions from simulation for the SM backgrounds are stacked
on top
of each other. The LM9 signal distribution is not stacked.
All distributions from simulation are normalized to the integrated luminosity of the data.}
\label{fig:MT2_MT2b_combinedHT}
\end{center}
\end{figure}
Figure~\ref{fig:MT2_MT2b_combinedHT} shows the \MTtwo distribution for events that satisfy the \MTtwob selection
criteria and with $\HT \geq 750\GeV$. As for the \MTtwo analysis
(Fig.~\ref{fig:MT2_combinedHT}),
the QCD multijet background
dominates for $\MTtwo<80\GeV$ but is strongly suppressed for $\MTtwo \geq 125\GeV$.
In the signal region, top-quark events dominate the electroweak contribution.
The white histogram (black dotted line) corresponds to the LM9 signal.
The corresponding event yields for data and SM simulation for the low and high \HT regions are summarized in Table~\ref{table:cut-flow-SMrelaxed}.
\begin{table}[htb!]
	\begin{center}
	\topcaption{
	Observed number of events and expected SM background event yields from simulation
	in the various \MTtwo bins for the \MTtwob event selection. These numbers are for guidance only and are not used
	in the final background prediction.}
	\label{table:cut-flow-SMrelaxed}
	\begin{tabular}{lcccc|cccccccccccc}
	\hline\hline
	 & QCD multijet& $W+$jets& Top & $Z(\nu\nu)+$jets& Total SM& Data \\ \hline \hline
		$750 \leq \HT < 950$ \\ \hline
	$\MTtwo [0,\infty]$  & 2.83e+04 & 4.53e+02 & 1.15e+03 & 1.41e+02 & 2.97e+04 & 2.99e+04 \\
	 $\MTtwo [125, 150]$ & 5.16 & 1.86 & 20.3 & 0.95 & 28.3 & 22 \\
	 $\MTtwo [150, 200]$ & 0.16 & 1.94 & 17.9 & 2.00 & 22.1 & 16 \\
	 $\MTtwo [200, 300]$ & 0.0 & 1.84 & 9.43 & 1.25 & 12.6 & 16 \\
	 $\MTtwo [300, \infty]$ & 0.0 & 0.57 & 2.55 & 0.53 & 3.65 & 2 \\
	\hline
		 $\HT \geq 950$ \\ \hline
	 $\MTtwo [0,\infty]$ & 1.19e+04 & 2.18e+01 & 5.46e+02 & 6.51e+00 & 1.25e+04 & 1.23e+04 \\
	 $\MTtwo [125, 150]$ & 1.25 & 0.76 & 9.95 & 0.64 & 12.7 & 10 \\
	 $\MTtwo [150, 180]$ & 0.57 & 0.79 & 7.15 & 0.43 & 8.96 & 10 \\
	 $\MTtwo [180, 260]$ & 0.67 & 1.09 & 6.62 & 0.68 & 9.06 & 9 \\
	 $\MTtwo [260, \infty]$ & 0.04 & 0.76 & 3.09 & 0.65 & 4.55 & 3 \\
	\hline\hline
	\end{tabular}
	\end{center}
\end{table}

\subsection{Background prediction and results}
\label{sec:bkgdrelaxed}

The QCD multijet contribution is estimated
following the same approach as for the \MTtwo analysis. We find that the
function in Eq.~\eqref{eq.bkgdpred.ratio} fitted to data in the region
$50<\MTtwo<80\GeV$ provides a good description of the QCD multijet background,
also for events containing b-tagged jets.
From the fit to data, the prediction of the QCD multijet background is obtained in the various \MTtwo bins for the low and high \HT regions.

Events arising from top-quark production are the dominant background
contribution in the signal region. The top-quark contribution is evaluated,
together with the one from W$(\ell\nu)$+jets, in the same way as for the \MTtwo
analysis,  using single-electron and single-muon events, as well as taus
decaying to hadrons.

The background from Z$(\cPgn\cPagn)$+jets events is expected to be very
small compared with the background from top-quark events.
We estimate the background from Z$(\cPgn\cPagn)$+jets events with the method based
on W+jets events discussed for the \MTtwo analysis.
As the selection of W$(\mu \nu)$+jets events includes a b-tag veto to suppress the top-quark background,
a ratio of efficiencies for W$(\mu \nu)$+jets events with a b tag to W$(\mu
\nu)$+jets events without a b tag is taken into account.
This ratio is obtained from simulation.

\begin{figure}[!htb]
  \begin{center}
  \includegraphics[width=0.49\textwidth]{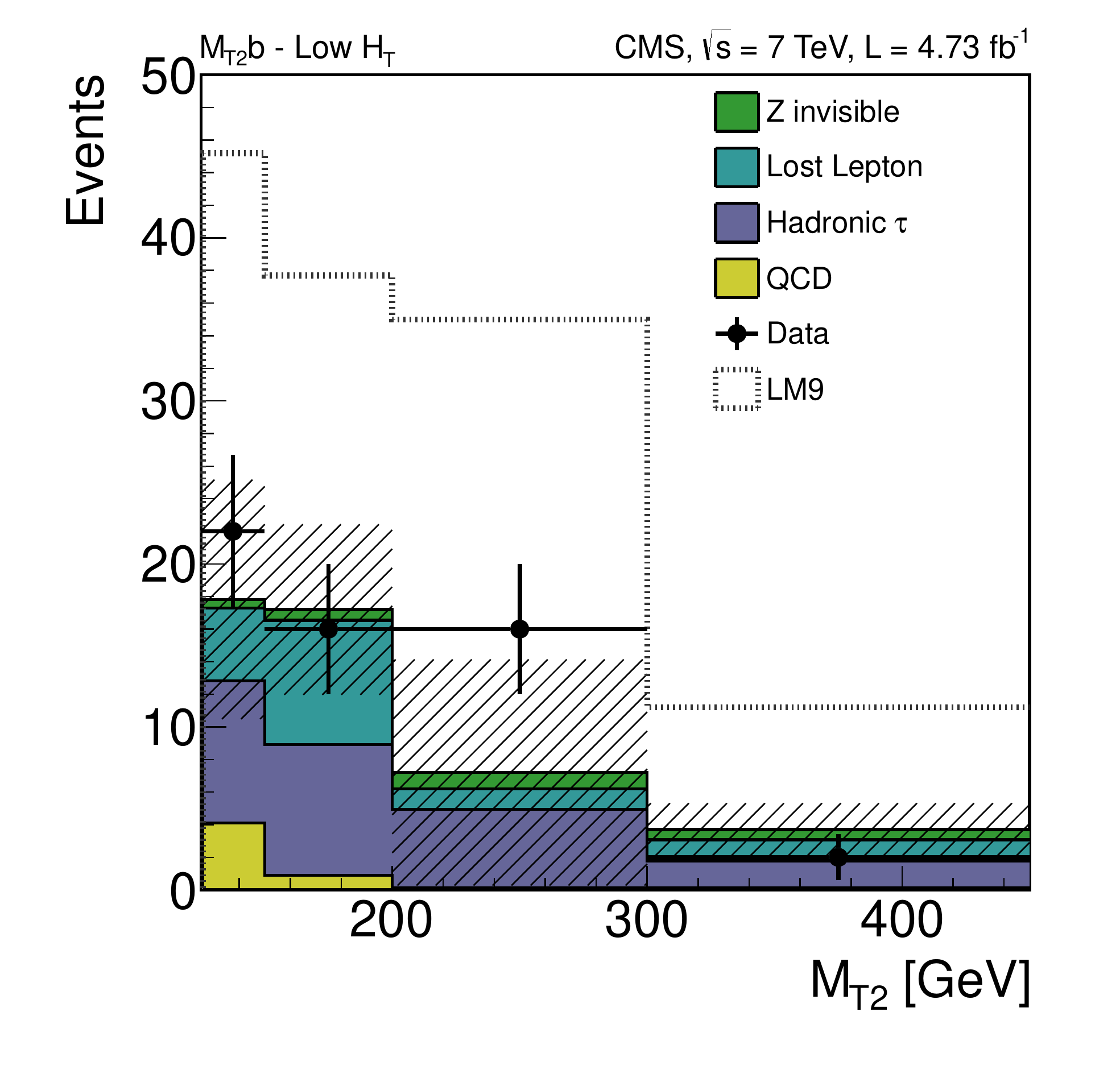}
  \includegraphics[width=0.49\textwidth]{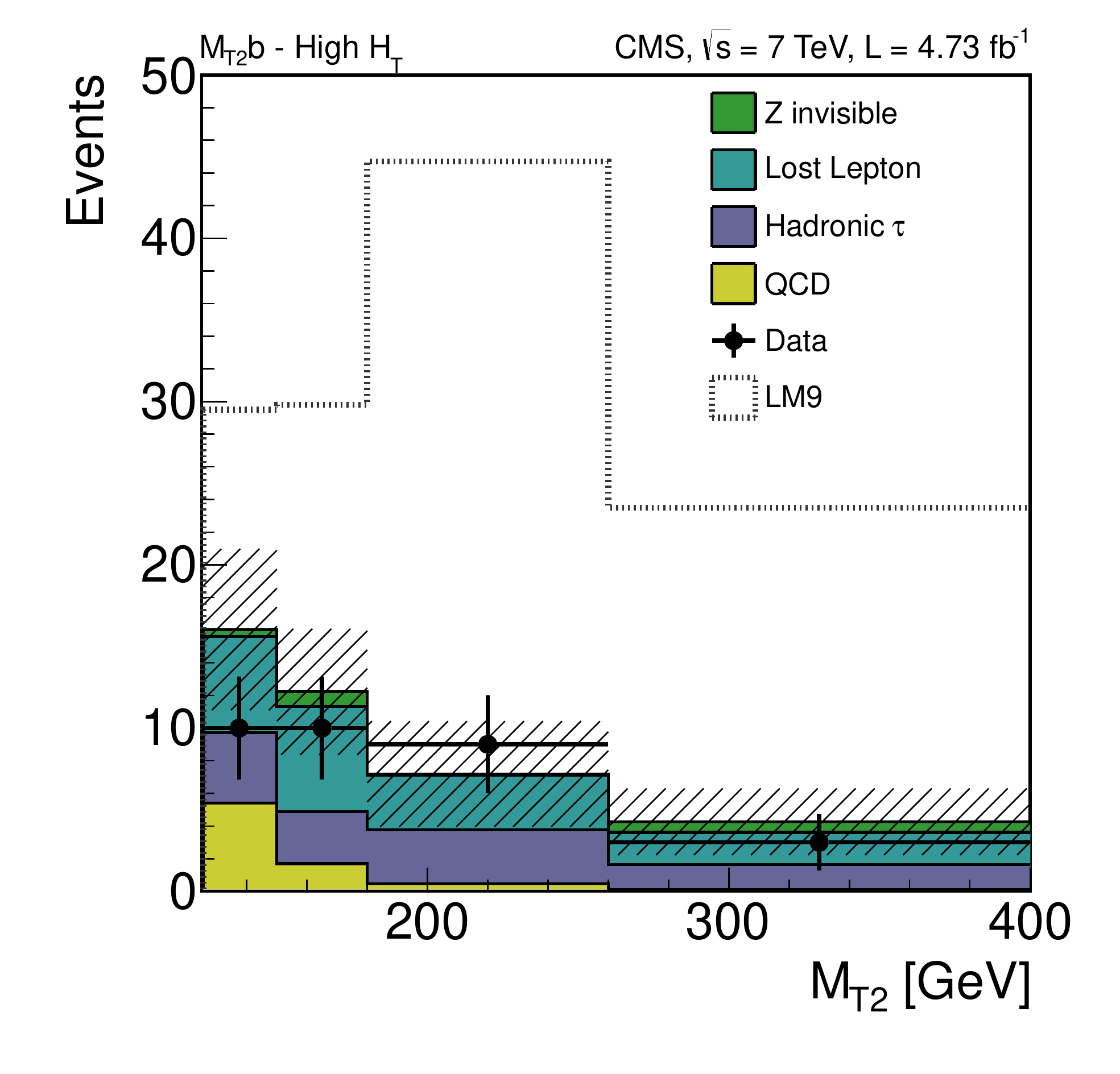}
  \caption{\MTtwo distribution from the background estimates compared to data for the \MTtwob selection.
	The figure on the left corresponds to the
    	$750\leq \HT<950$\GeV region, while that on the right corresponds to $\HT \geq 950\GeV$.
  The prediction from simulation for the LM9 signal model (not stacked) are also shown.
The hatched band shows the total uncertainty on the SM background estimate.}
  \label{fig:datadrivenMT2b}
  \end{center}
\end{figure}

\begin{table}[!htb]
	\begin{footnotesize}
	\begin{center}
	\topcaption{Estimated event yields for each background contribution in
        the various \MTtwo and \HT bins. The predictions from control regions
        in data are compared to the expected event yields from simulation.
        Statistical and systematic uncertainties are added in quadrature. The
        total background prediction is compared to data in the last two
        columns.}
	\label{table:datadrivenMT2b}
	\begin{tabular}{@{ }l|cc|cc|c|cc|c|c@{ }}
	\hline\hline
	& \multicolumn{2}{c|}{$Z\to\cPgn\cPagn$} & \multicolumn{2}{c|}{Lost lepton} & $\tau\to \text{had}$ & \multicolumn{2}{c|}{QCD multijet} & Total bkg. & Data \\
	            & sim. & data pred. & sim. & data pred. & Estimate & sim. & data pred. &  data pred. & \\\hline
	$750 \leq \HT < 950$ \\ \hline
	$\MTtwo[125,150]$       & 1.0 & $0.5\pm0.4$ & 12.8 & $4.5\pm3.2$ & $8.7\pm6.3$ & 5.16 & $4.1\pm2.1$    & $17.8\pm7.3$ & 22\\
	$\MTtwo[150,200]$       & 2.0 & $0.7\pm0.3$ & 11.3 & $7.6\pm3.6$ & $8.0\pm3.8$ & 0.16 & $0.90\pm0.51$  & $17.2\pm5.2$ & 16\\
	$\MTtwo[200,300]$       & 1.3 & $1.0\pm0.5$ & 6.1  & $1.3\pm1.7$ & $4.9\pm6.7$ & 0.0  & $0.04\pm0.03$  & $7.2\pm6.9$  & 16\\
	$\MTtwo[300, \infty]$   & 0.5 & $0.6\pm0.3$ & 1.3  & $1.3\pm0.9$ & $1.8\pm1.3$ & 0.0  & $0.00\pm0.00$  & $3.7\pm1.6$  & 2\\
	\hline
	      $\HT \geq 950$ \\ \hline
	$\MTtwo[125,150]$      & 0.6 & $0.4\pm0.3$ & 6.2 & $5.9\pm3.3$ & $4.3\pm2.4$ & 1.25 & $5.4\pm2.8$     & $16.0\pm4.9$ & 10\\
	$\MTtwo[150,180]$      & 0.4 & $0.9\pm0.4$ & 4.6 & $6.4\pm3.3$ & $3.2\pm1.7$ & 0.57 & $1.7\pm0.9$     & $12.2\pm3.9$ & 10\\
	$\MTtwo[180,260]$      & 0.6 & $0.1\pm0.1$ & 4.2 & $3.4\pm2.3$ & $3.3\pm2.3$ & 0.67 & $0.45\pm0.25$   & $7.2\pm3.2$  & 9\\
	$\MTtwo[260,\infty]$   & 0.6 & $0.7\pm0.4$ & 2.2 & $2.0\pm1.6$ & $1.6\pm1.3$ & 0.04 & $0.05\pm0.04$   & $4.3\pm2.0$  & 3\\
	\hline\hline
	\end{tabular}
	\end{center}
	\end{footnotesize}
\end{table}
The results of the estimates for the various backgrounds
are summarized in Table~\ref{table:datadrivenMT2b} and shown in Fig.~\ref{fig:datadrivenMT2b}.

\section{Statistical Interpretation of the results and exclusion limits}
\label{sec:exclusion}

No significant deviation from the SM background prediction is observed
and upper limits are set on a potential signal.
The statistical approach used to derive limits
follows closely the methodology of Ref.~\cite{Higgscombi11}.
A brief description of the steps relevant
to this analysis follows.

First, a likelihood function is constructed as the product of Poisson
probabilities for each \HT, \MTtwo search bin. These probabilities
are functions
of the predicted signal and background yields in each bin. Systematic uncertainties
are introduced as nuisance parameters in the signal and background models.
Log-normal distributions are taken as a suitable choice for the probability
density distributions for the nuisance parameters.

In order to compare the compatibility of the data with the background-only
and the signal-plus-background hypotheses, we construct the test statistic
$q_\lambda$ based on the profile likelihood ratio:
\begin{equation}
  \label{eq:testStat}
  q_\lambda=-2\ln \frac{\mathcal L (\text{data}|\lambda,\hat\theta_\lambda)}
{\mathcal L (\text{data}|\hat\lambda,\hat\theta)},\qquad
\text{with}\; 0\le\hat\lambda\le\lambda,
\end{equation}
where the signal strength modifier $\lambda$ is introduced to test signal cross
section values $\sigma=\lambda\sigma_\text{sig}$.
Both the denominator and the numerator are maximized.
In the numerator, the signal parameter strength $\lambda$ remains fixed and the
likelihood is maximized only for the nuisance parameters,
whose values at the maximum are denoted $\hat\theta_{\lambda}$.
In the denominator, the likelihood is maximized for both $\lambda$ and $\theta$.
$\hat\lambda$ and $\hat\theta$ denote the values
at which $\mathcal L$ reaches its global maximum in the denominator.
The lower constraint $0\le\hat\lambda$ is imposed because the signal strength
cannot be negative,
while the upper constraint $\hat{\lambda}<\lambda$ guarantees a one-sided
confidence interval.
The value of the test statistic for the actual observation
is denoted $q_\lambda^\text{obs}$.
This test statistic~\cite{Higgscombi11} differs from
that used at LEP and the Tevatron.

To set limits, a modified frequentist CL$_\mathrm{s}$ approach is
employed~\cite{Junk1999,Read2002}.
We first define the probabilities to obtain an outcome of an experiment
at least as signal-like as the one observed for the background-only
and for the signal-plus-background hypotheses. The CL$_\mathrm{s}$
quantity is then defined as the ratio of these two probabilities.
In the modified frequentist approach, the value of CL$_\mathrm{s}$ is
required to be less than or equal to $\alpha$ in order to establish a $(1-\alpha)$
confidence level (CL) exclusion. To quote the upper limit on $\lambda$ for a
given signal at 95\% CL, we adjust $\lambda$ until we reach CL$_\mathrm{s}$ = 0.05.

\subsection{Exclusion limits in the CMSSM plane}
\label{sec:exclusion.cmssm}

\begin{figure}[htb!]
	\begin{center}
	\includegraphics[width=0.8\textwidth]{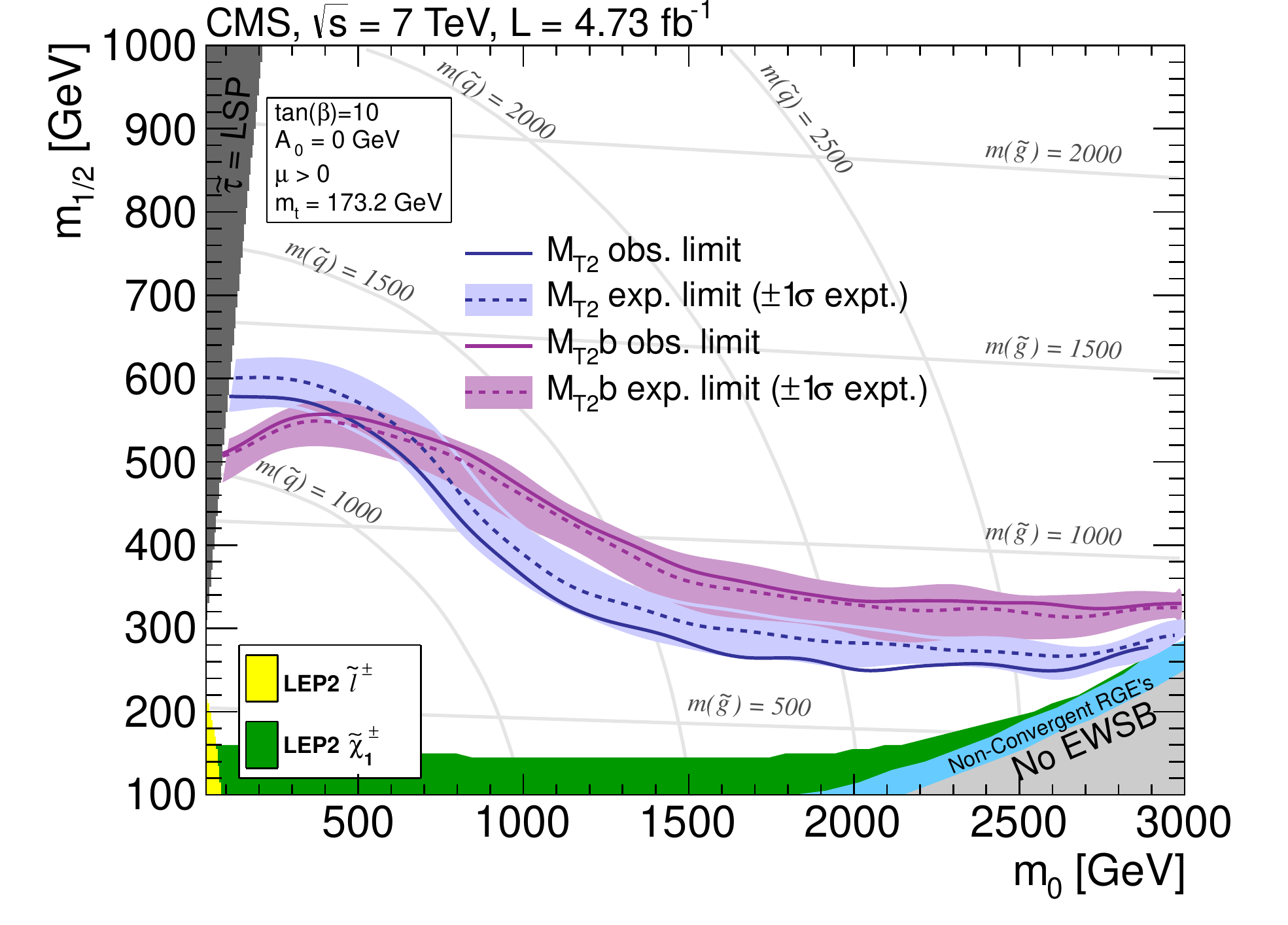}
	\includegraphics[width=0.8\textwidth]{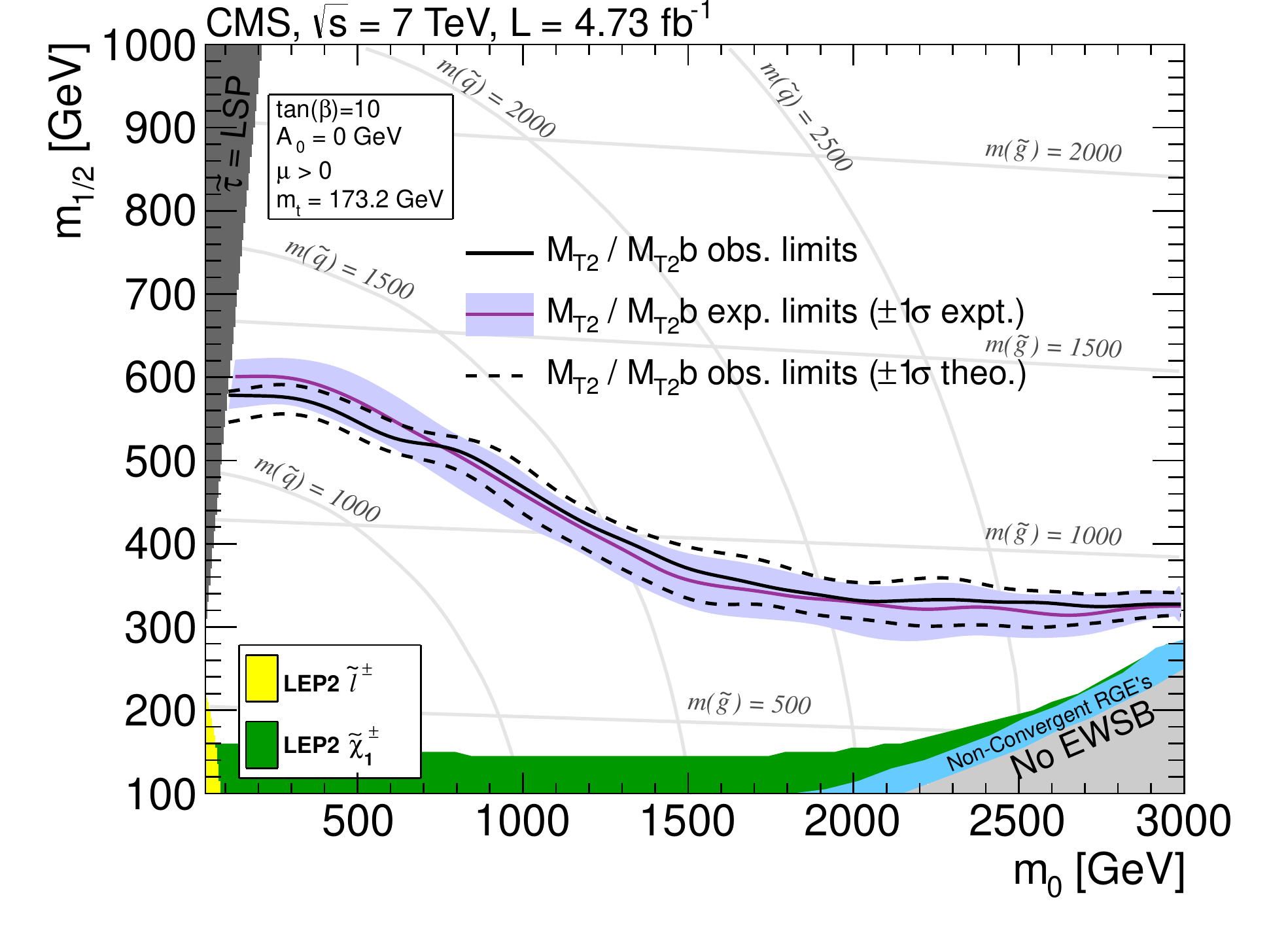}
	\caption{Top: exclusion limit in the CMSSM $(m_0, m_{1/2})$ plane for the \MTtwo and  \MTtwob analyses
		 with $\tan\beta = 10$. Bottom: Combined limit based on the best expected limit at each point.
}
\label{fig:exclm0m12}
\end{center}
\end{figure}

Exclusion limits at 95\% CL are determined in the CMSSM $(m_0,
m_{1/2})$ plane~\cite{Matchev:2012vf}. The signal cross section is calculated at NLO and
next-to-leading-log (NLL)
accuracy~\cite{Prospino97,Kulesza:2008jb,Kramer:2012bx}
At each point in the scan four CL$_{\rm s}$ values are computed for $\lambda =
1$:
the observed, the median expected, and the one standard deviation $(\pm 1
\sigma)$ expected bands.
If the corresponding CL$_{\rm s}$ value is smaller than 0.05, the point is excluded
at 95\% CL,
resulting in the exclusion limits shown in Fig.~\ref{fig:exclm0m12}.
The results from both the \MTtwo and \MTtwob selections
are shown in Fig.~\ref{fig:exclm0m12}~(top). In
Fig.~\ref{fig:exclm0m12}~(bottom), the results are combined into
a single limit exclusion curve based on the best expected limit at
each point of the plane.

The dominant sources of systematic uncertainties on the signal model
are found to be the jet energy scale and (for the \MTtwob analysis) the
b-tagging efficiency. These two
uncertainties are evaluated at each point of the CMSSM
plane, typically ranging from 5 to 25\% for the former and from 2 to 6\%
for the latter. Additionaly, a 2.2\% uncertainty is associated
with the luminosity determination~\cite{CMS-PAS-SMP-12-008}. All these
uncertainties are included in the statistical interpretation as nuisance
parameters on the signal model.

Observed exclusion limits are also determined
when the signal cross section is varied by changing the renormalization
and factorization scales by a factor of 2 and using the PDF4LHC
recommendation~\cite{pdf4lhc} for the PDF uncertainty. The exclusion contours
obtained from this method are shown by the dashed curves of
Fig.~\ref{fig:exclm0m12} and referred to as theory uncertainties.

The effect of signal contamination in the leptonic control region
could be significant, yielding a potential background overprediction
of about 1-15\%. To account for this effect, the signal yields are corrected
by subtracting the expected increase in the background estimate that would
occur if the given signal were present in the data.

The results in Fig.~\ref{fig:exclm0m12} (top) establish that the \MTtwo
analysis is powerful in the region of large squark and gluino masses, corresponding to small $m_0$ and large $m_{1/2}$,
while the \MTtwob analysis increases sensitivity to large squark and small
gluino masses,  corresponding to large $m_0$ and small $m_{1/2}$.
Conservatively, using the minus one standard deviation $(-1\sigma)$ theory uncertainty
values of the observed limit,
we derive absolute lower limits on the squark and gluino masses for the chosen CMSSM parameter set.
We find lower limits of $m(\sQua) > 1110$\GeV and $m(\sGlu) > 800$\GeV,
as well as $m(\sQua) = m(\sGlu)  > 1180$\GeV assuming equal squark and gluino masses.

\subsection{Exclusion limits for simplified model spectra}
\label{sec:exclusion.simple}

In this section we interpret the results in terms of
simplified model spectra~\cite{SMS1},
which allow a presentation of the exclusion potential in the context
of a larger variety of
fundamental models, not necessarily
in a supersymmetric framework. We studied the following topologies:
\begin{itemize}
 \item gluino pair production, with $\sGlu \to  \cPq \cPaq \chiz$;
 \item gluino pair production, with $\sGlu \to  \cPqb \cPaqb \chiz$;
 \item gluino pair production, with $\sGlu \to  \cPqt \cPaqt \chiz$;
 \item gluino pair production, with $\sGlu \to  \cPq \cPaq \cPZ \chiz$.
\end{itemize}
The last of these models is used to demonstrate the sensitivity of the analysis in a high jet multiplicity topology, since the hadronic decay of the $\cPZ$ boson
can lead to (maximally) 8 jets in the final state.
\begin{figure}[!t]
\begin{center}
\includegraphics[width=0.49\textwidth]{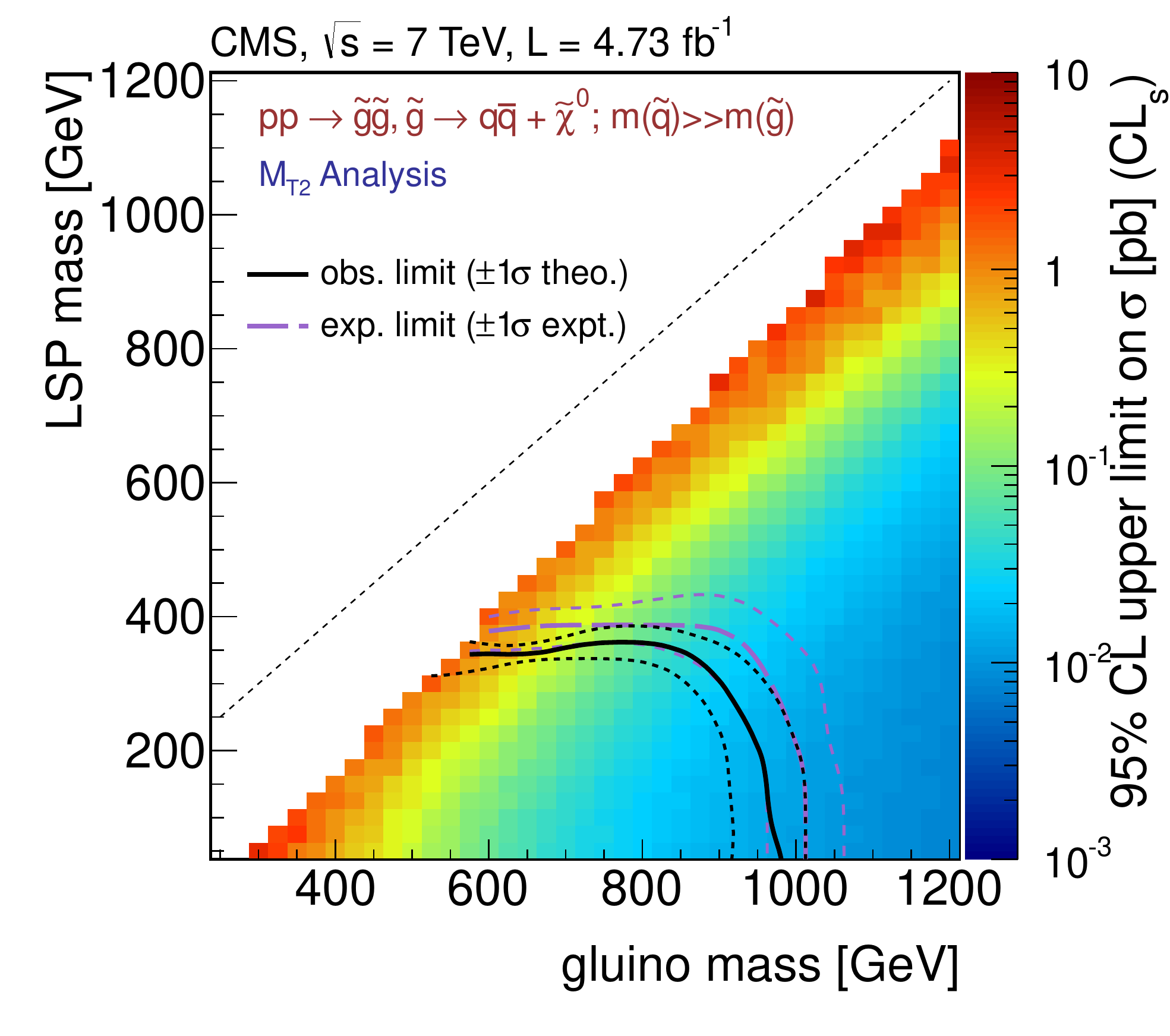}
\includegraphics[width=0.49\textwidth]{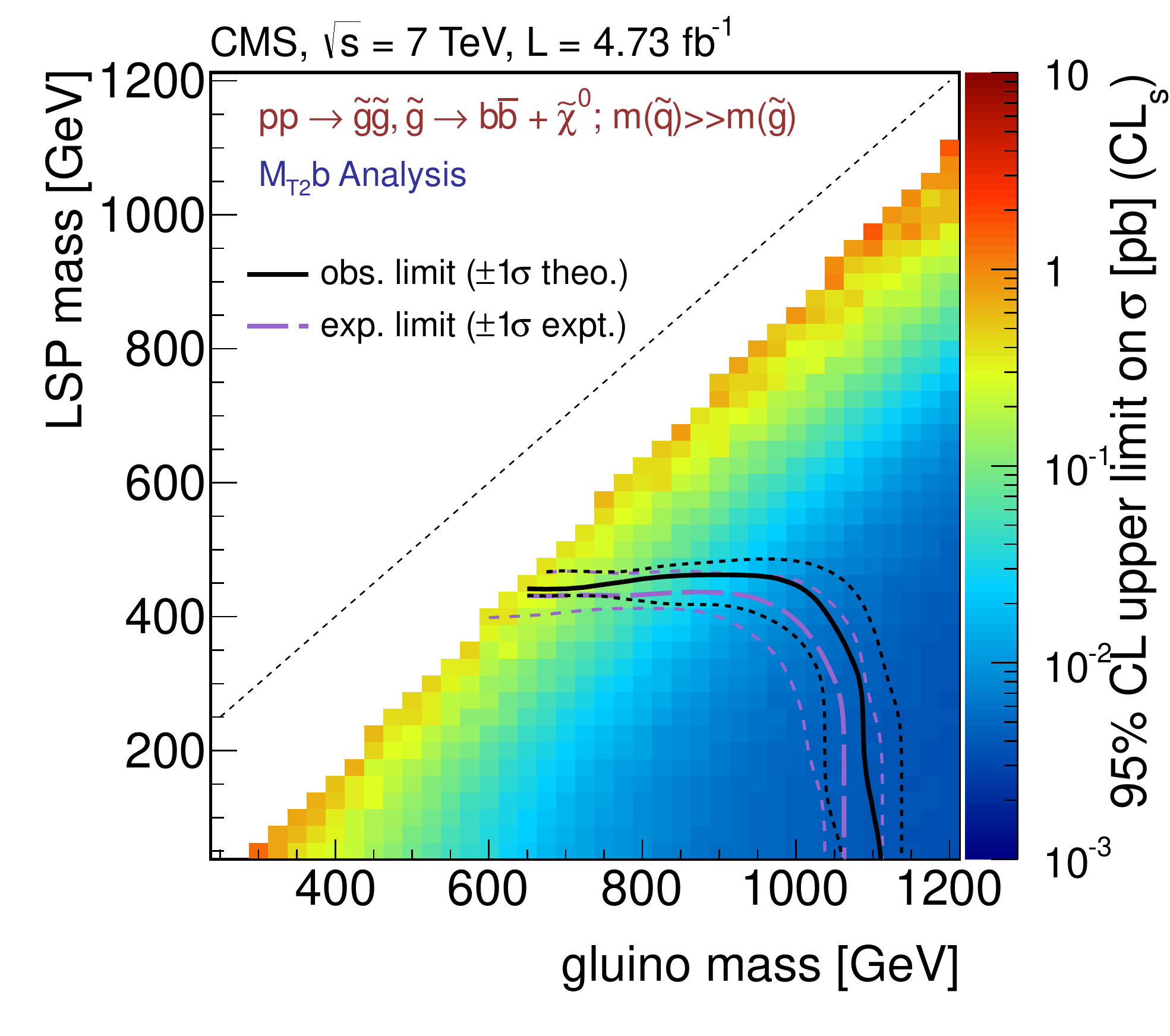}
\includegraphics[width=0.49\textwidth]{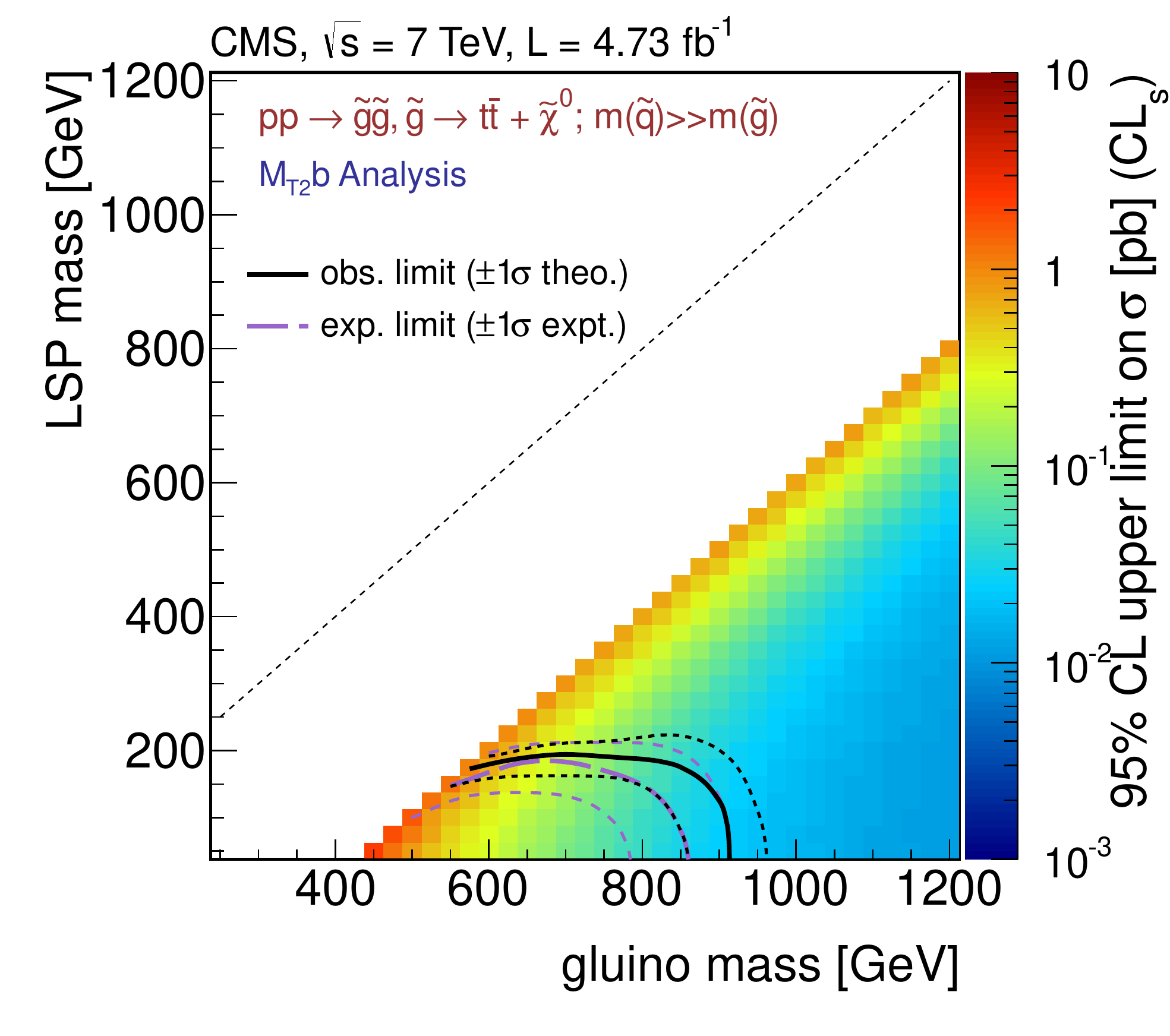}
\includegraphics[width=0.49\textwidth]{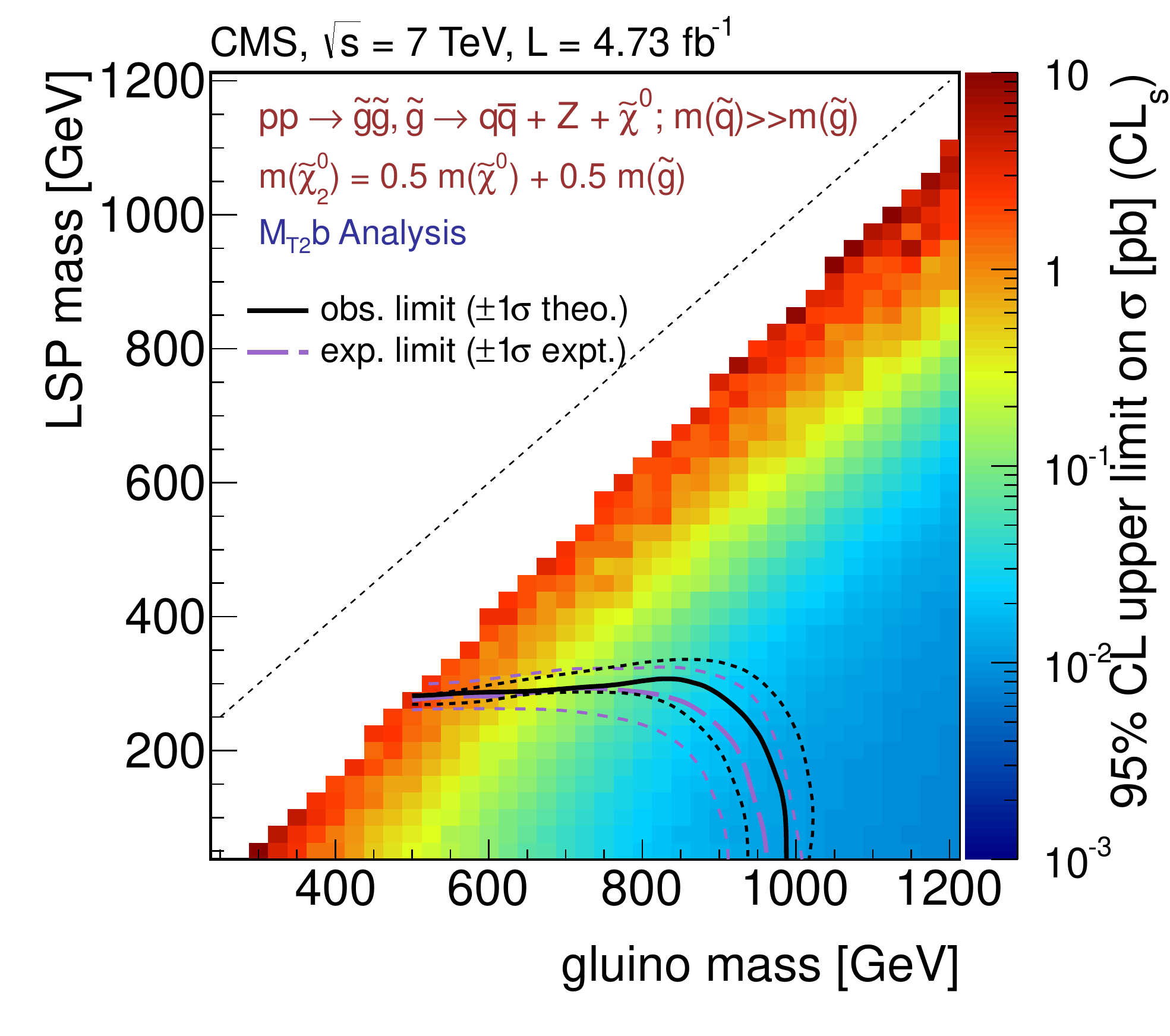}
\caption{Exclusion limits for simplified model spectra. Upper left:  gluino pair
production with $\sGlu \to \cPq \cPaq \chiz$ using the \MTtwo analysis. Upper
right: gluino pair production with $\sGlu \to \cPqb \cPaqb \chiz$, using the
\MTtwob analysis. Lower left: gluino pair production with $\sGlu \to \cPqt \cPaqt \chiz$, using the \MTtwob analysis.
Lower right: gluino pair production with $\sGlu \to \cPq \cPaq \cPZ \chiz$,
using the \MTtwob analysis. The signal production cross sections are
calculated at NLO and NLL accuracy~\cite{Prospino97,Kulesza:2008jb,Kramer:2012bx}.}
\label{fig:exclSMS_T1bbbb_MT2b}
\end{center}
\end{figure}
In Fig.~\ref{fig:exclSMS_T1bbbb_MT2b} the 95\% CL excluded cross sections
are reported as a function of the relevant masses for gluino pair
production with $\sGlu \to \cPq \cPaq \chiz$ using the \MTtwo
analysis, and for $\sGlu \to \cPqb \cPaqb \chiz$, $\sGlu \to \cPqt \cPaqt \chiz$
and $\sGlu \to \cPq \cPaq \cPZ \chiz$ using the \MTtwob analysis. Systematic
uncertainties on jet energy scale and on b-tagging efficiencies are taken
into account as nuisance parameters on the signal model.
To minimize the effect of ISR modeling uncertainties, the region near the
diagonal is excluded in the limit setting.
Observed, median expected, and one standard deviation ($\pm1\sigma$
experimental) expected limit curves are derived for the
nominal signal cross section. Also shown are
the $\pm1\sigma$ variation in the observed limit when the signal cross
section is varied by its theoretical uncertainties.

\section{Summary}
\label{sec:conclusion}

We have conducted a search for supersymmetry or similar new physics in
hadronic final states using the \MTtwo variable calculated from massless
pseudojets.
\MTtwo is strongly correlated with \MET for SUSY processes, yet provides a
natural suppression of QCD multijet background.
The data set for this analysis corresponds to \Lumino of integrated luminosity
in  $\sqrt{s} = 7$\TeV pp collisions
collected with the CMS detector during the 2011 LHC run.
All candidate events are selected using hadronic triggers.
Two complementary analyses are performed.
The \MTtwo analysis targets decays of moderately heavy squarks and gluinos,
which naturally feature a sizeable \MET.
This analysis is based on events containing three or more jets and no isolated
leptons. We show that the tail of the \MTtwo distribution, obtained after this
selection, is sensitive to a potential SUSY signal. A second approach, the
\MTtwob analysis, is designed to increase the sensitivity to events with heavy
squarks and light gluinos, in which the \MET tends to be smaller.
Therefore, the restriction on \MTtwo  is relaxed. The effect of the loosened
\MTtwo is compensated by requiring at least one b-tagged jet and a larger jet
multiplicity, to suppress the QCD multijet background.
For both analyses, the standard model backgrounds, arising from QCD
multijet, electroweak, and top-quark production processes, are
obtained from data control samples and simulation.
No excess beyond the standard model expectations is found. Exclusion limits
are established in the CMSSM parameter space, as well as for some simplified
model spectra.
Conservatively, using the minus one standard deviation $(-1\sigma)$ theory
uncertainty values, absolute mass limits in the CMSSM scenario for
$\tan\beta = 10$ are found to be $m(\sQua) > 1110\GeV$ and $m(\sGlu) >
800\GeV$, and $m(\sQua) = m(\sGlu) > 1180$\GeV assuming equal squark and
gluino masses.

\input{acknow}

%% file: acknow.tex
\section*{Acknowledgements}

We congratulate our colleagues in the CERN accelerator departments for the
excellent performance of the LHC machine. We thank the technical and
administrative staff at CERN and other CMS institutes, and acknowledge support
from: FMSR (Austria); FNRS and FWO (Belgium); CNPq, CAPES, FAPERJ, and FAPESP
(Brazil); MES (Bulgaria); CERN; CAS, MoST, and NSFC (China); COLCIENCIAS
(Colombia); MSES (Croatia); RPF (Cyprus); MoER, SF0690030s09 and ERDF
(Estonia); Academy of Finland, MEC, and HIP (Finland); CEA and CNRS/IN2P3
(France); BMBF, DFG, and HGF (Germany); GSRT (Greece); OTKA and NKTH
(Hungary); DAE and DST (India); IPM (Iran); SFI (Ireland); INFN (Italy); NRF
and WCU (Korea); LAS (Lithuania); CINVESTAV, CONACYT, SEP, and UASLP-FAI
(Mexico); MSI (New Zealand); PAEC (Pakistan); MSHE and NSC (Poland); FCT
(Portugal); JINR (Armenia, Belarus, Georgia, Ukraine, Uzbekistan); MON,
RosAtom, RAS and RFBR (Russia); MSTD (Serbia); SEIDI and CPAN (Spain); Swiss
Funding Agencies (Switzerland); NSC (Taipei); TUBITAK and TAEK (Turkey); STFC
(United Kingdom); DOE and NSF (USA).  

Individuals have received support from the Marie-Curie programme and the
European Research Council (European Union); the Leventis Foundation; the
A. P. Sloan Foundation; the Alexander von Humboldt Foundation; the Belgian
Federal Science Policy Office; the Fonds pour la Formation \`a la Recherche
dans l'Industrie et dans l'Agriculture (FRIA-Belgium); the Agentschap voor
Innovatie door Wetenschap en Technologie (IWT-Belgium); the Council of Science
and Industrial Research, India; the Iran National Science Foundation (INSF); 
the Compagnia di San Paolo (Torino); and the HOMING PLUS programme of Foundation 
for Polish Science, cofinanced from the European Union, Regional Development Fund.

%% file: SUS-12-002-authorlist.tex
\textbf{Yerevan Physics Institute,  Yerevan,  Armenia}\\*[0pt]
S.~Chatrchyan, V.~Khachatryan, A.M.~Sirunyan, A.~Tumasyan
\vskip\cmsinstskip
\textbf{Institut f\"{u}r Hochenergiephysik der OeAW,  Wien,  Austria}\\*[0pt]
W.~Adam, T.~Bergauer, M.~Dragicevic, J.~Er\"{o}, C.~Fabjan\cmsAuthorMark{1}, M.~Friedl, R.~Fr\"{u}hwirth\cmsAuthorMark{1}, V.M.~Ghete, J.~Hammer, N.~H\"{o}rmann, J.~Hrubec, M.~Jeitler\cmsAuthorMark{1}, W.~Kiesenhofer, V.~Kn\"{u}nz, M.~Krammer\cmsAuthorMark{1}, D.~Liko, I.~Mikulec, M.~Pernicka$^{\textrm{\dag}}$, B.~Rahbaran, C.~Rohringer, H.~Rohringer, R.~Sch\"{o}fbeck, J.~Strauss, A.~Taurok, P.~Wagner, W.~Waltenberger, G.~Walzel, E.~Widl, C.-E.~Wulz\cmsAuthorMark{1}
% ...
\vskip\cmsinstskip
\dag:~Deceased\\
1:~~Also at Vienna University of Technology, Vienna, Austria\\
% ...
